\documentclass[pdftex,12pt]{article}

\newcommand{\narrow}[1]{}  

\usepackage{amsmath,texdef2004,latexsym,amsfonts}
\usepackage{cite}
\usepackage[pdftex]{graphicx}
\DeclareGraphicsExtensions{.pdf}
\usepackage{color}

\newcommand{\expansion}[2][\pm1]{\langle #2\rangle_{\set{#1}}}

\newcommand{\Rplot}[1]{\begin{picture}(50,35)
\put(5,2){\makebox(0,0)[bl]{\includegraphics{#1}}}
\put(0,20){$R_2$}
\put(30,0){$R_1$}\end{picture}}

\margin
\date{April 20, 2009}
\author{David Tse\thanks{D.~Tse (email:  {\em dtse@eecs.berkeley.edu})
   is  with Wireless Foundations, EECS Dept.,  University of California
    at Berkeley and acknowledges the support of NSF grant CNS-0722032.}%
 \and Roy Yates
\thanks{R.~Yates (email: {\em ryates@winlab.rutgers.edu}) is with
  WINLAB,   E\&CE Dept., Rutgers University and acknowledges the support of NSF
  grant CNS-0721826.}
}

\newcommand{\entropyhat}[2][d]{\hat{H}_{#1}(#2)}
\newcommand{\nmax}{n_{\max}}
\newcommand{\rate}[2][i]{r_{#2}^{(#1)}}
\newcommand{\levelgap}[2][i]{\delta_{#2}^{(#1)}}
\newcommand{\levelsum}[1][i]{\delta^{(#1)}}
\newcommand{\pcross}{p}

\newcommand{\Renh}{\overline{R}}
\newcommand{\Senh}{\overline{S}}
\newcommand{\pc}[1]{\prob{C(#1)}}

\newcommand{\epshat}{\hat{\epsilon}}
\newcommand{\re}[1]{{#1}_I}
\newcommand{\im}[1]{{#1}_Q}

\renewcommand{\ccdf}[2]{\overline{F}_{#1}(#2)}
\newcommand{\mc}{\mbox{---}}
\newcommand{\FF}{\mbox{$\mathbb{F}$}}

\renewcommand{\var}[1]{\text{var}[#1]}

\newcommand{\B}[2][\w]{\beta_{#1}(#2)}
\newcommand{\Bt}[2][\w]{\tilde{\beta}_{#1}(#2)}

\newcommand{\Cinc}[1]{\Delta_C(#1)}

\newcommand{\mmse}[1]{\text{mmse}\paren{#1}}
\newcommand{\C}{\mathcal C}
\newcommand{\Ystate}[1]{Y^{(#1)}}

\newcommand{\Ys}{\Ystate{s}}

\newcommand{\Rcal}{{\mathcal R}}
\newcommand{\Rcaltil}{\tilde{\Rcal}}
\newcommand{\Pcal}{{\mathcal P}}
\newcommand{\Xcal}{{\mathcal X}}

\newcommand{\Ical}{{\mathcal I}}
\newcommand{\Ncal}{{\mathcal N}}

\newcommand{\Iset}{\Ical}
\newcommand{\Nset}{\Ncal}
\newcommand{\Itilset}{\tilde{\Ical}}
\newcommand{\stge}{\ge_{\text{st}}}

\newcommand{\limiter}[2][1]{\min\bracket{#1,#2}}

\newcommand{\CN}{\mathcal{CN}}

\newcommand{\interior}{\text{int}}

\newcounter{lettercount}
\newenvironment{letterate}{\begin{list}%
{(\alph{lettercount})}{\usecounter{lettercount}}}{\end{list}}

\begin{document}
\title{{\Large\bf Fading Broadcast Channels with State Information at
    the Receivers}}
\maketitle
  \begin{abstract}
Despite considerable progress on the information-theoretic broadcast
channel, the capacity region of fading broadcast channels with
channel state known at the receivers but unknown at the transmitter
remains unresolved. We address this subject by introducing a layered
erasure broadcast channel model in which each component channel has
a state that specifies the received signal levels in an instance of
a deterministic binary expansion channel. We find the capacity
region of this class of broadcast channels. The capacity achieving
strategy assigns each signal level to the user that derives the
maximum expected rate from that level.  The outer bound is based on a
channel enhancement
that creates a degraded broadcast channel for which the capacity
region is known. This same approach is then used to find inner and
outer bounds to the capacity region of fading Gaussian broadcast
channels.
The achievability scheme employs a superposition of binary
inputs. For intermittent AWGN channels and for Rayleigh
fading channels, the achievable rates are observed to be with 1-2 bits
of the outer bound at high SNR. We also prove that the achievable rate
region is within 6.386 bits/s/Hz of the capacity region for all fading
AWGN broadcast channels.
\end{abstract}
\section{Introduction}
The two most basic multiuser communication scenarios are many-to-one
and one-to-many, captured by the multiple access channel (MAC) and
the broadcast channel (BC) respectively. While the capacity region
of the general multiple access channel is known since the 70's
\cite{Ahl71,Liao-thesis72}, that of the general broadcast channel
is still open. Yet, progress has been made in the past 40 years on
special cases.  A class of channels, of particular importance to
wireless communication, are the Gaussian broadcast channels (AWGN
BC). The capacity region of AWGN BC's is known, for both the case
when the channel state is fixed and time-invariant, and for the case
when the channel-state is time-varying (fading). This is true for
single antenna \cite{Cover:72,Ber74,Hug75,Tse97,LG2001} or for
multiple antenna channels \cite{WSS06}.  The key assumption behind
these results is that the channel state is known at the transmitter
as well as the receivers (perfect CSI). However, the problem becomes
open once the assumption of CSI at the transmitter is removed, even
when the transmitter and each of the receivers has only a single
antenna. This is an important scenario in practice since in a fast
fading environment, it may be difficult to feedback channel state
information in a timely fashion to the transmitter from the
receivers. Moreover, most cellular systems operate on a
frequency-division duplex (FDD) mode rather than on a time-division
duplex (TDD) mode, so the downlink channel information cannot be
inferred from uplink channel measurements. Nevetheless, despite its
apparent practical importance, there are very few results on the
capacity of the fading BC with receiver-only CSI (see for example
\cite{TuninettiShamai2003,JafarianVishwanath2008}).


The channels mentioned above for which the capacity region is known
are either degraded (in the case of the single-antenna
time-invariant channel), parallel with degraded components
(time-varying single-antenna channels with perfect CSI) or have a
related degraded structure (MIMO BC's). The fading broadcast
channel with only receiver CSI has no such degraded structure for
arbitrary fading distributions, thus making it a challenging problem
from a theoretical standpoint.

In this paper, we focus on the simplest scenario with two receivers
and a single antenna at the transmitter and at each of the
receivers. Our main contribution is two-fold:

\begin{itemize}

\item We propose a {\em layered erasure broadcast channel} to
approximate the Gaussian fading channel and determine its capacity
region exactly. The erasures in this model are {\em correlated},
and, like the Gaussian fading BC, the layered erasure BC is neither
degraded nor parallel with degraded components.

\item Using the insights from the erasure model, we derive a new
outer bound to the Gaussian fading BC capacity region and
demonstrate a binary expansion superposition (BES) scheme that
achieves rates within $6.386$ bits/s/Hz per user to the outer bound.
This gap holds in the worst case over {\em all} fading
distributions. We also demonstrate example fading distributions for
which the gap is much smaller.

\end{itemize}

The layered erasure BC is based on a new point-to-point erasure
channel model. This model provides a simpler way of thinking about
fading and may be of independent interest. The transmitted signal is
thought of as a vector of bits, from the most significant to less
significant bits. The bits can be viewed as layers of the
transmitted signal. Fading is modeled as erasures of the less
significant bits, and how many bits are erased depends on the
instantaneous channel strength. Erasures are correlated because when
a bit is erased, all the less significant bits are also erased. The
layered erasure model can be thought of as a time-varying version of
the binary expansion deterministic channel model introduced by
Avestimehr, Diggavi and Tse \cite{ADT-deterministic-allerton}. While
modeling fading as erasures has appeared in the literature (see for
example \cite{Gupta-erasure2006}), typically these models regard the
{\em entire}\emph{} transmitted signal as erased and thus cannot
capture the continuous nature of the channel strength in the
Gaussian model. We do note, however, that this fading model has
appeared in the control literature \cite{Martins2006}.

\section{Background and Definitions}\label{sect:background}
As introduced by Cover \cite{Cover:72}, the two-user memoryless
broadcast channel (BC) consisting of a transmitter with input $X$
and receiver observations $Y_1$ and $Y_2$ described by a channel
transition probability $\pmf{Y_1,Y_2|X}{y_1,y_2|x}$. Through this
multiuser channel, the  sender wishes to communicate private messages
at rate $R_i$ to receiver $i$ as well as a common message at rate
$R_0$ to both receivers.



In this work, we focus on the $R_0=0$ case where there is no common
message. Even in this case, the general BC capacity region is
unknown. However, the capacity region of the important special case of
the degraded channel is known. A broadcast channel $P_{Y_1,Y_2|X}$ is
{\em degraded} if there exists a
Markov chain $X\mc Y_1\mc Y_2$  that yields the
 marginal conditional distributions $\pmf{Y_1|X}{y_1|x}$ and
 $\pmf{Y_2|X}{y_2|x}$ consistent with $P_{Y_1,Y_2|X}$.
The capacity region $\Rcal$ of the degraded memoryless BC
\cite{Gallager-degraded,Ber74} is given in
the following theorem which we restate
here.
\begin{theorem}\thmlabel{KM5}
The capacity region $\Rcal$ of the
degraded memoryless BC $P_{Y_1,Y_2|X}$ is the union over all
$V,X$ such that $V\mc X\mc Y_1Y_2$ of rate pairs $(R_1,R_2)$ satisfying
\begin{subequations}
\begin{align}
R_1&\le I(X;Y_1|V),\\
R_2&\le I(V;Y_2).
\end{align}
\end{subequations}
\end{theorem}

To characterize the boundary of the capacity region $\C$ of a particular BC, we define  the weighted
sum rate maximization problem
\begin{equation}
\max_{(R_1,R_2)\in\C} \w_1R_1+\w_2R_2.
\eqnlabel{Rstar}
\end{equation}
To find the rate region $\C$,
we define $\omega=\omega_2/\omega_1$ and we solve
\begin{equation}\eqnlabel{Rstar2}
R^*(\w)=\max_{(R_1,R_2)\in\C} R_1+\omega R_2
\end{equation}
For each $\w$, the solution will be  associated with a pair
$(R^*_1,R^*_2)$ that
defines a capacity region constraint
\begin{equation}\eqnlabel{extreme-pt-constraints}
R_1+\w R_2 \le R^*_1+\w R^*_2,\qquad (R_1,R_2)\in \C.
\end{equation}
At $\w=0$, we obtain $(R^*_1,R^*_2)=(C_1,0)$ and the constraint
$R_1\le C_1$ where $C_1$ is the
ergodic capacity of the fading channel to receiver~$1$. Similarly, as
$\w\goes\infty$, we obtain the corner point $(R^*_1,R^*_2)=(0,C_2)$
corresponding to the ergodic capacity constraint on $R_2$.
In general, however, we will obtain an outer bound such that the
constraint \eqnref{extreme-pt-constraints} may not be tight. 

\begin{figure}
\setlength{\unitlength}{0.9mm}\thicklines
\begin{center}
\begin{picture}(100,50)(-10,-10)
\put(0,0){\vector(1,0){85}}
\put(0,0){\vector(0,1){35}}
\put(0,35){\makebox(0,0)[b]{$R_2$}}
\put(85,0){\makebox(0,0)[l]{$R_1$}}
\put(-1,25){\makebox(0,0)[r]{$C_2$}}
\put(75,-1){\makebox(0,0)[t]{$C_1$}}
\qbezier(0,25)(69,24)(75,0)
%
\put(30,0){\line(0,1){1}}
\put(30,-2){\makebox(0,0)[t]{$1$}}
\put(0,7.5){\line(1,0){1}}
\put(-1,7.5){\makebox(0,0)[r]{$\omega_1$}}
\put(0,20){\line(1,0){1}}
\put(-1,20){\makebox(0,0)[r]{$\omega_2$}}
\put(0,0){\vector(4,1){30}}
\qbezier[40](0,0)(20,5)(70.59,17.65)
\qbezier[6](70.59,17.65)(71,16)(72.8,8.8)
\put(72.8,8.8){\line(1,-4){2.2}}
\put(81,4){\makebox(0,0)[cl]{\bf 1}}
\put(80,4){\vector(-1,0){6}}
\put(0,0){\vector(3,2){30}}
\qbezier[40](0,0)(30,20)(54.46,36.31)
\qbezier[25](54.46,36.31)(58,31)(65,20.5) 
\put(65,20.5){\line(2,-3){7.8}}
\put(74,16){\makebox(0,0)[cl]{\bf 2}}
\put(74,16){\vector(-1,0){6}}
\qbezier[20](0,0)(5,20)(8.65,34.59)
\qbezier[40](8.65,34.59)(26,30.25)(47,25)
\put(47,25){\line(4,-1){18}}
\put(0,25){\makebox(0,0){$\bullet$}}%
\put(0,25){\line(1,0){47}}%
\put(47,25){\makebox(0,0){$\bullet$}}%
\put(65,20.5){\makebox(0,0){$\bullet$}}%
\put(72.8,8.8){\makebox(0,0){$\bullet$}}%
\put(75,8){\makebox(0,0)[bl]{$(R_1^*,R_2^*)$}}
\put(75,0){\makebox(0,0){$\bullet$}}%
\end{picture}
\end{center}
\caption{In the outer bound to a capacity region, each extreme point,
  denoted by $\bullet$, of an outer bound region is
  specified by a pair of constraints. In this example, the boundary
  segments marked {\bf 1} and {\bf 2} correspond to the constraint
  \protect\eqnref{extreme-pt-constraints} with $\w=\w_1$ and $\w=\w_2$
  respectively.}
\label{fig:extreme-points}
\end{figure}
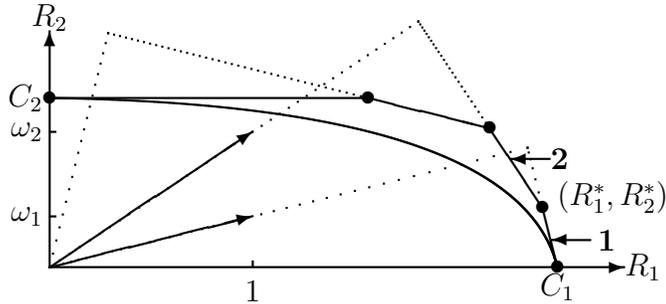

When the channel state distributions are
discrete, the point $(R^*_1,R^*_2)$ generally will be associated with an
interval $[\w_1,\w_2]$ of weights and a family of constraints
\begin{equation}
R_1+\w R_2 \le R_1^*+\w R_2^*,\qquad (R_1,R_2)\in \C, \w\in[\w_1,\w_2].
\end{equation}
In this case, as shown in Figure~\ref{fig:extreme-points}, the endpoint
weights $\w_1$ and $\w_2$ define  a pair of constraints such that
$(R^*_1,R^*_2)$ is an extreme point of an outer bound region.
The outer bound region will be the convex hull of the set of extreme
points and the origin $(0,0)$.

In our subsequent derivations, we assume $\w\ge 1$; i.e.,
bits communicated to receiver $2$ are favored. For the case $\w<1$,
we factor $\w$ out of \eqnref{Rstar2} and we see that the boundary of
the capacity region for $\w<1$ is given by the optimization problem
\begin{equation}\eqnlabel{Rstar2reverse}
R^*(\w)= \w \max_{(R_1,R_2)\in\Rcal} (1/\w)R_1+ R_2.
\end{equation}
In this case, we have the identical optimization as in \eqnref{Rstar2}
but with the roles of users $1$ and $2$ reversed in that we now favor
receiver~1 by the weight $1/\w$. However, as the labeling of receivers
$1$ and $2$ is arbitrary, we can simply reverse label the receivers
and solve \eqnref{Rstar2} with $1/\w\ge 1$.  Henceforth, our
derivations using the weighted sum rate
maximization \eqnref{Rstar2} will assume a weight $\w\ge 1$. We will
see that the solution will remain valid for $\w<1$.

In this paper, we adopt several conventions. For a random variable
$N$, the probability mass function (PMF) is $\pmf{N}{n}:=\prob{N=n}$
and the complementary cumulative distribution function (CDF) is
$\ccdf{N}{n}:=\prob{N\ge n}$. The ${\cal N}(0,1)$ random variable $Z$
has complementary CDF $Q(z)=\prob{Z\ge z}$. We define $\floor{x}$
as the largest integer less than or equal $x$, $\ceiling{x}$ as the
smallest integer greater than or equal to $x$, and $x^+:=\max(0,x)$.
We also define $\sgn(x)$ as the sign of $x$. That is, $\sgn(x)=-1$ for
$x<0$ and $\sgn(x)=1$ for $x\ge 0$. All logarithms are to the base 2
unless otherwise noted. Proofs 
appear in the appendix.


\section{Layered Erasure Broadcast Channel}\label{sect:layered-erasure}
\subsection{Channel Model}
We start by reviewing the broadcast channel formulation of the
binary expansion deterministic channel model of
\cite{ADT-deterministic-allerton}.
Each communication channel from the sender  to a receiver $j$  is
associated with a non-negative integer gain $n_{j}$ that describes
how many signal ``levels'' are observed at receiver $j$. The best
channel in the system supports $q=\max_{j} n_{j}$ levels. At each
time $t$, the sender  transmits a vector
$\Xv_i([t])\in\FF_2^q$. Algebraic definition of the received signals
is based on the $q\times q$ ``shift'' matrix
\begin{equation}
\Smat=\begin{bmatrix}
0 & 0 & 0 & \cdots & 0\\
1 & 0 & 0 & \cdots & 0\\
0 & 1 & 0 & \cdots & 0\\
\vdots & \ddots & \ddots & \ddots &\vdots\\
0 & \cdots & 0 & 1 & 0
\end{bmatrix}.
\end{equation}
For example, if $\Yv=\Smat^2\Xv$, we have that $Y_1=Y_2=0$, and
 $Y_k=X_{k-2}$ for $k=3,\ldots,q$. We note that $\Smat^0$ is the $q\times
 q$ identity matrix.
In terms of $\Smat$,  the received signals are
\begin{align}
\Yv_i[t]&= \Smat^{q-n_{i}}\Xv[t],\qquad i=1,2,
\end{align}
where summation and multiplication are over the binary field
$\FF_2$. Please refer to
\cite{ADT-deterministic-allerton} for further details.

The number of levels $n_i$ observed
by receiver $i$ is intended to describe the SNR of the communication
channel. We model a fading channel by replacing $n_i$ by a
non-negative random variable $N_i$ such that $0\le N_i\le q$. The
channel state at receiver $i$ is given by
$\set{N_i[t]|t=1,2,\ldots}$,  an iid random sequence with PMF
$\pmf{N_i}{n}$.   We will assume receiver channel state information
(CSI) in that $N_i[t]$ is known to receiver $i$ at time $t$.

As the channel states are iid, and the channels conditioned on the
channel state are memoryless, we drop the symbol time index $t$ for
convenience. When
the transmitter signals $\Xv=\tvec{X_1 & X_2 & \cdots & X_q}$,
receiver $i$ observes
\begin{align}
\Yv_i&=\Smat^{q-N_{i}}\Xv\\
&= \tvec{0 & \cdots& 0& X_1 & X_2 & \cdots &  X_{N_i}}.
\end{align}
Since receiver $i$ knows the channel
state $N_i$,  receiver $i$ knows
that the first $q-N_i$ zeroes in $\Yv_i$ carry no
data and that the data carrying signals are given by
\begin{equation}
(Y_{i,q-N_i+1},\ldots,Y_{q})=(X_1,\ldots,X_{N_i}).
\end{equation}
The missing signal components $X_{N_i+1},\ldots,X_q$ have been
erased by the fading channel. For convenience, we use $X^n$ to denote
the signal vector $(X_1,\ldots,X_n)$. Thus the
transmitted signal is $X=X^q$ and when the channel state at
receiver $i$ is $N_i$, the receiver observation is
\begin{equation}
Y_i=X^{N_i}=(X_1,\ldots,X_{N_i}).
\end{equation}
\begin{definition}
A $q$-bit layered erasure channel  has input $X=X^q\in\FF_2^q$, and
output $Y=X^N$ where $N$ is an integer channel state that is
independent of $X^q$  satisfying $\prob{N\ge0}=1$ and $\prob{N\ge
q+1}=0$.
\end{definition}
Some useful properties of the $q$-bit layered erasure channel are
gathered in the following lemma.
\begin{lemma}\label{lem:qbit-fade}\label{lem:HIsum}
For a $q$-bit layered erasure channel with output $X^N$ and Markov
chain $V\mc X^q\mc X^N$,
\begin{letterate}
\item $\displaystyle I(X^q;X^N|V) =H(X^N|V,N)$,

\item $\displaystyle H(X^N|V,N)=\sum_{n=1}^q
\ccdf{N}{n}H(X_n|X^{n-1},V)$,
\item $\displaystyle I(V;X^N)= \sum_{n=1}^q\ccdf{N}{n}
I(V;X_n|X^{n-1})$
\end{letterate}
\end{lemma}
The proof appears in the Appendix.
Note that for a trivial $V$, Lemma~\ref{lem:qbit-fade}(a) implies
$I(X^q;X^N)=H(X^N|N)$.

We observe that the channel state PMF $\ipmf{N}$ completely
specifies a $q$-bit layered erasure channel. Given channel state
PMFs $\pmf{N_i}{n}$, the broadcast channel with input $X=X^q$ and
receiver observations $Y_1=X^{N_1}$ and $Y_2=X^{N_2}$ is described
by a pair of transition probability matrices $F_{N_1}$ and
$F_{N_2}$. In the parlance of \cite{Marton77},  a $q$-bit layered
erasure broadcast channel is simply a discrete memoryless BC
$(F_{N_1},F_{N_2})$. For convenience, we denote this broadcast channel
simply by the tuple $(N_1,N_2)$ of channel fading random variables.

In the sequel, we will exploit the structure of the degraded
broadcast channel. In the context of $q$-bit layered erasure
channels, the following definition and lemma show that
degradedness is associated with stochastic dominance of the $q$-bit
fading distributions.
\begin{definition}
Random variable $N_1$ is {\em stochastically larger} than
$N_2$, denoted $N_1\stge N_2$, if $\ccdf{N_1}{x}\ge \ccdf{N_2}{x}$
for all $x\ge 0$.
\end{definition}

\begin{lemma}  The $q$-bit layered erasure broadcast
  channel $(N_1,N_2)$ satisfying $N_1\stge N_2$ is a   degraded
  broadcast channel.
\end{lemma}


\subsection{Layered erasure  broadcast channel capacity}
To identify the boundary of the capacity region $\C$ of the
$q$-bit layered erasure broadcast channel, we start with the weighted
sum rate maximization problem \eqnref{Rstar2}.
To describe $\C$, we define the function
\begin{equation}
\B{n} :=\ccdf{N_1}{n} -\w\ccdf{N_2}{n}
\end{equation}
and construct the partition
$\Iset_1(\w)\cup\Iset_2(\w)$ where
\begin{subequations}\eqnlabel{Iset1}
\begin{align}
\Iset_1(\w)&=\set{n|\B{n}>0},\\
\Iset_2(\w)&=\set{n|\B{n}\le 0}.
\end{align}
\end{subequations}
With these definitions, we can state the capacity region $\C$.
\begin{theorem}\thmlabel{layered-erasure-capacity}
The capacity region $\C$ of the $q$ bit  layered erasure broadcast
channel $(N_1,N_2)$ is the convex hull of the union over all $\w\ge0$
of rate pairs $(R_1,R_2)$ satisfying
\begin{subequations}
\eqnlabel{opt-rate-i}
\begin{align}
R_1&\le \sum_{n\in \Iset_1(\omega)}\ccdf{N_1}{n},\\
R_2&\le\sum_{n\in \Iset_2(\omega)}\ccdf{N_2}{n}.
\end{align}
\end{subequations}
\end{theorem}
For each $\w\ge 0$, we present in Section~\ref{sect:binary-achievable} a
simple scheme that achieves any weighted sum rate on the boundary of
$\C$.  In
Section~\ref{sect:binary-converse}, we then use the
method of channel enhancement to create a degraded
broadcast channel that provides a matching outer bound.

\subsection{Layered Erasure BC: Achievability}
\label{sect:binary-achievable}
We assign signal levels $n\in \Iset_i(\w)$ to user $i$. In addition, we
employ independent signaling on each level. That is,  $\seq{X}{1}{q}$
are iid Bernoulli $(p=1/2$) random variables in each symbol period and
inputs $\set{X_n|n\in \Iset_i(\omega)}$ are used to communicate with
receiver $i$.

If level $n$ is assigned to user $i$,
the level $n$ output at receiver $i$ is erased if $N_i<n$. That is, the
erasure probability on the level $n$ subchannel is
$1-\ccdf{N_i}{n}$. With independent coding on each level, the erasure
channel at level $n\in \Iset_i(\w)$ enables
reliable communication with receiver $i$ at rate
$\ccdf{N_i}{n}$.
Thus
\begin{align}
\Rhat_i(\w)&=\sum_{n\in \Iset_i(\omega)}\ccdf{N_i}{n},\qquad i=1,2,
\end{align}
is the expected number of bits addressed and delivered without
erasure to receiver $i$ per symbol period.
Thus this scheme enables reliable communication to the receivers at
the weighted sum rate
\begin{equation}
\Rhat(\w)=\Rhat_1(\w)+\w \Rhat_2(\w).\eqnlabel{Rhat-rate}
\end{equation}
We note that assignment of level $n$ bits to user $1$ would contribute
$\ccdf{N_1}{n}$ to the weighted sum rate objective while the
alternative assignment of those same bits to user $2$ would contribute
$\w\ccdf{N_2}{n}$ to the objective. Hence our achievability scheme is
nothing more than a simple greedy assignment policy. While this
strategy is simple, it is not at all clear whether it is optimal. This
policy employs independent communication on subchannel corresponding
to different levels even though erasures at each receiver are
correlated across these subchannels.

\subsection{Layered Erasure BC: Converse}\label{sect:binary-converse}
To verify the outer bound to the weighted sum rate, we start with the
case $\w>1$ in which we favor communication with
receiver $2$. Later we will return to examine the case $\omega<1$.

Our outer bound will employ a degraded broadcast channel.
We note that for arbitrary $N_1$ and $N_2$, the $q$-bit layered
erasure broadcast channel is not  degraded. For $\w>1$, receiver $1$
is the less favored receiver.  Our approach will be to enhance the
channel to this less-favored receiver by replacing the fading
distribution $N_1$ by $\Ntil_1$ such that $\Ntil_1\stge N_1$. It
follows that any reliable broadcast communication strategy that
communicates at rates $(R_1,R_2)$ through the $q$-bit layered
erasure BC $(N_1,N_2)$ is also a reliable strategy through the
$q$-bit fading BC $(\Ntil_1,N_2)$, implying $\Rcal(N_1,N_2)\subseteq
\Rcal(\Ntil_1,N_2)$. We will show for each $\w$ that the
enhancement of the channel for the less favored receiver will not
result in the transmission of additional information through that
enhanced channel. Consequently, the maximum weighted sum rate
$R_1+\w R_2$ will be the same whether the maximization is over
$\Rcal(N_1,N_2)$ or $\Rcal(\Ntil_1,N_2)$.

The enhanced channel
for  receiver 1 is given by
\begin{equation}
\ccdf{\Ntil_1}{n}=\limiter{\max(\ccdf{N_1}{n},\w\ccdf{N_2}{n})}.
\eqnlabel{N1enhanced}
\end{equation}
Similar to the achievability scheme which was defined in terms of
$\B{n}$,  the
converse will employ the weighted difference of complementary CDFs
\begin{equation}\eqnlabel{Bt-defn}
\Bt{n} := \ccdf{\Ntil_1}{n} - \w \ccdf{N_2}{n}.
\end{equation}
We note that the channel enhancement
implies that
\begin{equation}\eqnlabel{Ntil-defn2}
\ccdf{\Ntil_1}{n}=\begin{cases}
\ccdf{N_1}{n}, & \B{n}>0,\\
\limiter{\w\ccdf{N_2}{n}}, & \B{n}\le 0.
\end{cases}
\end{equation}
Taken together, \eqnref{Bt-defn} and \eqnref{Ntil-defn2} imply the
following claim.
\begin{lemma}\label{lem:Isets}
\begin{equation}
\Iset_1(\w)=\set{n|\B{n}>0} =\set{n|\Bt{n}>0}.
\end{equation}
For $n\in\Iset_1(\w)$, $\ccdf{\Ntil_1}{n}=\ccdf{N_1}{n}$ and
$\Bt{n}=\B{n}$.
\end{lemma}

Before proceeding, we note that it is easy to verify that $\Ntil_1$ is
a well-defined $q$-bit fading random variable. Next we observe that
$\ccdf{\Ntil_1}{n}\ge \ccdf{N_1}{n}$ and
thus $\Ntil_1\stge N_1$.
Finally, we observe that one can verify that $\Ntil_1\stge N_2$.
It then follows that $(\Ntil_1,N_2)$ is a degraded $q$-bit broadcast
channel.

We now use Lemma~\ref{lem:qbit-fade} to apply \Thmref{KM5} to the
$q$-bit layered erasure broadcast channel $(\Ntil_1,N_2)$.  The
capacity region $\Rcaltil=\Rcal(\Ntil_1,N_2)$ is the set of all rate
pairs $(\Rtil_1,\Rtil_2)$ satisfying
\begin{subequations}
\eqnlabel{Vregion}
\begin{align}
\Rtil_1&\le H(X^{\Ntil_1}|\Ntil_1,V),\\
\Rtil_2&\le I(V;X^{N_2})
\end{align}
\end{subequations}
for some $(V,X^q,X^{\Ntil_1},X^{N_2})\in\Pcal(\Ntil_1,N_2)$.
We see  that the weighted sum rate achieved by any feasible rate pair
$(R_1,R_2)\in\Rcaltil$ is upper bounded by the weighted sum rate
\begin{equation}
\Rtil^*(\w,V)=\Rtil^*_1(V)+\w \Rtil^*_2(V)
\end{equation}
associated with the corner rate pair
 \begin{align}
 \Rtil^*_{1}(V)&=H(X^{\Ntil_1}|\Ntil_1,V), &
\Rtil^*_{2}(V)&=I(V;X^{N_2}),
 \eqnlabel{RBcorner}
 \end{align}
for some auxiliary $V$. We now identify the
auxiliary $V$ that maximizes $\Rtil^*(\w,V)$.
Applying Lemma~\ref{lem:qbit-fade}, we obtain
the weighted rate
\begin{align}
\Rtil^*(\w,V)&= H(X^{\Ntil_1}|\Ntil_1,V)+\omega I(V;X^{N_2})\\
&=\sum_{n=1}^q\ccdf{\Ntil_1}{n}H(X_n|X^{n-1},V)
\narrow{\nonumber\\&\qquad}
+\sum_{n=1}^q\omega\ccdf{N_2}{n}I(V;X_n|X^{n-1})\\
&=\sum_{n=1}^q\Bt{n}H(X_n|X^{n-1},V)
\narrow{\nonumber\\&\qquad}
+\w\sum_{n=1}^q \ccdf{N_2}{n}H(X_n|X^{n-1}).
\eqnlabel{RBsum}
\end{align}
We observe that:
\begin{itemize}
\item  The first sum in \eqnref{RBsum} is maximized if
\begin{equation}
H(X_n|X^{n-1},V)= \begin{cases}
1 & \Bt{n}>0,\\
0 & \Bt{n}\le 0.
\end{cases}
\eqnlabel{HgivenV}
\end{equation}
\item The second sum in \eqnref{RBsum} is
maximized by choosing the $\seq{X}{1}{q}$ to be iid
Bernoulli $(p=1/2)$ random variables.
\end{itemize}
However, these two requirements
are not contradictory. Given that we choose the $X_i$ to be iid
Bernoulli $(p=1/2)$ random variables, we can meet the requirement
\eqnref{HgivenV} by
choosing
\begin{equation}
V=\Vtil=\set{X_n|\Bt{n}\le 0}
\end{equation}
In particular, we note that the role of the
auxiliary is to carry the information for the favored receiver whose bits are
given greater weight.

Applying the optimal  $\Vtil$  to \eqnref{RBsum}, we see
that $R^*(\w)$, the maximum weighted sum
rate over all feasible rate pairs $(R_1,R_2)\in\Rcal(N_1,N_2)$, is
upper bounded by
\begin{equation}
\Rtil^*(\w,\Vtil)= \sum_{n: \Bt{n}>0}\Bt{n}+\w\sum_{n=1}^q
\ccdf{N_2}{n}.\eqnlabel{Rtil-rate}
\end{equation}
Applying Lemma~\ref{lem:Isets}, we obtain
\begin{align}
\Rtil^*(\w,\Vtil)&= \sum_{n\in\Iset_1(\w)}\ccdf{N_1}{n}
+\w\sum_{n\in\Iset_2(\w)}\ccdf{N_2}{n},
\eqnlabel{Rtil-final}
\end{align}
which is identical to $\Rhat(\w)$, the achievable weighted sum rate
\eqnref{Rhat-rate}. We note that this tight outer bound can also be
obtained by the K\"orner-Marton outer bound in \cite{Marton77}.

We now return to examine the outer bound for weight $\w <1$.
As discussed in Section~\ref{sect:background}, we can repeat the steps
corresponding to equations~\eqnref{Vregion}
through~\eqnref{Rtil-final} with labels $1$ and $2$ reversed and $\w$
replaced by $1/\w$. Under this role reversal, we now enhance
the channel to the less-favored receiver $2$
by replacing the fading
distribution $N_2$ by $\Ntil_2$ so that the
$q$-bit layered erasure BC $(N_1,\Ntil_2)$ has a degraded receiver
$1$. The maximization of the weighted sum rate for the degraded
 memoryless BC $(N_1,\Ntil_2)$ yields the same requirements as before:
the inputs $X_i$
 must be iid Bernoulli $(p=1/2)$ random variables and the auxiliary
 $V$ specifies the bits destined for the favored receiver, which is
 now receiver~$1$.
The corresponding bit assignment rule, derived from \eqnref{Iset1}
with labels $1$ and $2$ reversed and $\w$ replaced by $1/\w$,
 becomes
\begin{align}
\Iset_1'(\omega) &=
\set{n|\ccdf{N_2}{n}\le\frac{1}{\w}\ccdf{N_1}{n}}.
\end{align}
However, trivial manipulation shows that
$\Iset_1'(\w)=\Iset_1(\w)$. That is, the optimal decision rule for allocating
bits to each user to maximize the weighted sum rate is unchanged by
the role reversal. Thus the resulting optimal weighted sum rate is still given
by the achievable rate in \eqnref{opt-rate-i} for all $\w>0$.

\subsection{Discussion}
\Thmref{layered-erasure-capacity} implies that the boundary of
the capacity region of the $q$ bit layered erasure broadcast channel
is defined by a finite set of points. In particular at $\omega=0$,
or equivalently $\omega_1>0$ and $\omega_2=0$, all bits can be
assigned to receiver $1$ and we obtain the extreme point
$\Rv^{(0)}=(R^{(0)}_1,R^{(0)}_2)$ such that
\begin{align}
R_1^{(0)}&=\sum_{j=1}^q \prob{N_1\ge j} =\E{N_1},
& R_2^{(0)}&=0.
\eqnlabel{extreme-pt0}
\end{align}
In addition, there is a collection of critical points
$\set{\seq{\omega}{1}{q}}$ such that
\begin{equation}
\prob{N_1\ge j} =\omega_j\prob{N_2\ge j},\qquad j=1,\ldots,q.
\end{equation}
We define $\w_0'=0$ and
$\set{\omega'_1,\omega'_2,\ldots,\omega'_{q'}}$
as the subset of unique $\omega_j$ arranged in strictly increasing
order. With these definitions, we define the set of closed intervals
$\set{\Omega_j|j=0,\ldots,q'}$ such that
\begin{equation}
\Omega_j=\begin{cases} [\omega'_j,\omega'_{j+1}]{} & 0\le j< q',\\{}
[\omega'_{q'},\infty) & j=q'.
\end{cases}
\end{equation}
Using $\interior(\Omega_J)$ to denote the interior of $\Omega_j$,
it follows that for all $\omega\in\interior(\Omega_j)$, $\Iset_1(\omega)$ is
unchanging; we denote this set by $\Iset_{1,j}$ and its complement by
$\Iset_{2,j}$. For any $\omega\in\interior(\Omega_j)$, the optimal solution to
the weighted sum rate maximization problem \eqnref{Rstar} is given
by assigning signaling bits in $\Iset_{i,j}$ to user $i$.
 This solution yields an extreme point, denoted
$\Rv^{(j)}=(R_1^{(j)},R_2^{(j)})$, of the rate region $\Rcal$.  When
$\w=\w_j'$ for some $j<q'$, the weighted sum maximization is
degenerate in 
that both extreme points $\Rv^{(j)}$ and $\Rv^{(j+1)}$ achieve maximum
weighted sum rate.
The full set of extreme points $\set{\Rv^{(0)},\ldots,\Rv^{(q')}}$
defines the boundary of the rate region. The entire region can then be
achieved by time-sharing among these extreme points.

\begin{table}
\begin{displaymath}
\begin{array}{cc}
\begin{array}[t]{cccc}
j & \prob{N_1\ge j} & \prob{N_2\ge j} & \omega_j\\\hline
1& 3/4 & 1/2 & 3/2\\
2 & 1/4 & 1/2 & 1/2
\end{array}
&
\begin{array}[t]{ccccc}
j & \Omega_j & \Iset_{1,j}& \Iset_{2,j} & (R_1^{(j)},R_2^{(j)})\\\hline
0& [0,1/2] & \set{1,2} & \phi & (1,0)\\
1& [1/2,3/2] &\set{1} & \set{2} & (3/4,1/2)\\
2& [3/2,\infty)& \phi&\set{1,2} & (0,1)
\end{array}\\
\text{\bf (a)} & \text{\bf (b)}
\end{array}
\end{displaymath}
\caption{Capacity region construction  for the example in
  \Eqnref{ErasureExample1}.}
\label{tab:example1}
\end{table}

\begin{table}
\begin{displaymath}
\begin{array}{cc}
\begin{array}[t]{cccc}
j & \prob{N_1\ge j} & \prob{N_2\ge j} & \omega_j\\\hline
1& 3/4 & 1/2 & 3/2\\
2 & 0 & 1/2 & 0
\end{array}
&
\begin{array}[t]{ccccc}
j & \Omega_j & \Iset_{1,j}& \Iset_{2,j} & (R_1^{(j)},R_2^{(j)})\\\hline
0& [0,0] & \set{1,2} & \phi & (3/4,0)\\
1& [0,1/2] &\set{1} & \set{2} & (3/4,1/2)\\
2& [3/2,\infty)& \phi&\set{1,2} & (0,1)
\end{array}\\
\text{\bf (a)} & \text{\bf (b)}
\end{array}
\end{displaymath}
\caption{Capacity region construction  for the example in
  \Eqnref{ErasureExample2}.}
\label{tab:example2}
\end{table}

\begin{figure}
\begin{center}
\begin{tabular}{cc}
\includegraphics{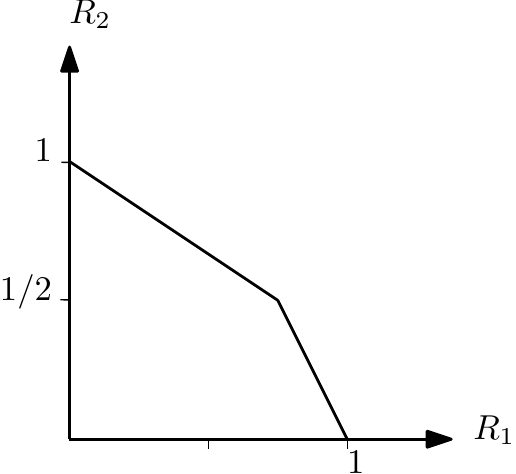} &
\includegraphics{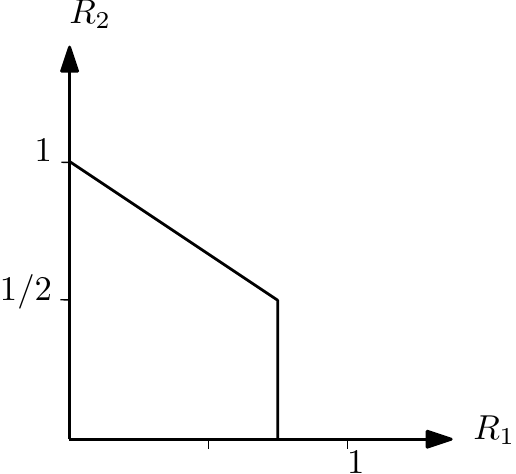}\\
{\bf (a)} & {\bf (b)}
\end{tabular}
\end{center}
\caption{Examples of $2$-bit capacity regions.}
\label{fig:BC-examples}
\end{figure}

To make this clear, consider the following example
\begin{equation}\eqnlabel{ErasureExample1}
\pmf{N_1}{n}=\begin{cases}
1/4, & n=0,\\
1/2, & n=1,\\
1/4, & n=2,
\end{cases}
\qquad
\pmf{N_2}{n}=\begin{cases}
1/2, & n=0,\\
0,  & n=1,\\
1/2, & n=2,
\end{cases}
\end{equation}
in which receiver $1$ has a more reliable look at the level $1$ bit
while receiver $2$ has a better look at the level $2$ bit. As a
consequence, it can be shown that  this $q$-bit fading channel
$(N_1,N_2)$ is not degraded, not less noisy, nor more capable, nor
is it semi-deterministic.  Since erasures at receiver $2$ are
correlated, this channel is not  a parallel channel with reversely
degraded components. Nevertheless, the capacity region of this
channel is easy to find. From the PMFs $\pmf{N_1}{n_1}$ and
$\pmf{N_2}{n_2}$, we construct the table shown in
Table~\ref{tab:example1}(a). This reveals
$\set{\w_1',\w_2'}=\set{1/2,3/2}$. The corresponding bit intervals,
bit assignments and extreme points are shown in
Table~\ref{tab:example1}(b). The resulting rate region is shown in
Figure~\ref{fig:BC-examples}(a).

Here is a second example with
\begin{equation}\eqnlabel{ErasureExample2}
\pmf{N_1}{n}=\begin{cases}
1/4 & n=0,\\
3/4 & n=1,\\
0    & n=2,
\end{cases}
\qquad
\pmf{N_2}{n}=\begin{cases}
1/2 & n=0,\\
0  & n=1,\\
1/2 & n=2.
\end{cases}
\end{equation}
From these PMFs, we construct Table~\ref{tab:example2}(a), which
yields $\set{\w_1',\w_2'}=\set{0, 3/2}$ and the intervals, bit
assignments and extreme points shown in Table~\ref{tab:example2}(b).
The resulting rate region is shown in Figure~\ref{fig:BC-examples}(b)
where we note that in the vertical boundary by the $R_1$ axis arises
because receiver $1$ never observes bit $X_2$ and thus suffers no
penalty when it is used by receiver $2$.
\section{Fading Gaussian BC}
\label{sect:fading-Gaussian}
Now  we draw an analogy
between  fading Gaussian broadcast channels  and the
layered erasure broadcast channel and use the analogy to derive good
schemes and a good outer bound for the capacity region of the fading
Gaussian broadcast channel.
We will show that the gap between the achievable rates of the
scheme and the outer bounds are within $6.386$
bits/s/Hz per user of each other, irrespective of the fading
processes.

For the Gaussian BC, consider fading processes such that
$(S_i,\theta_i)$ is the channel state of
receiver $i$, where $S_i$ is a real-valued non-negative channel gain and
$\theta_i \in [0, 2\pi]$ is the channel phase.  Random
variables $S_i, \theta_i$ are assumed to be independent of each other
and interpreted to be independent from symbol time to symbol time. The
received signal of user $i$ at a particular time is
\begin{equation}\eqnlabel{FBC-complex}
\Ytil_i = \sqrt{S_i} e^{j\theta_i} \Xtil + \Ztil_i,\qquad i=1,2,
\end{equation}
where $\Xtil$ is the complex baseband transmitted signal with unit power
constraint and $\Ztil_i \sim \CN(0,1)$.  The channel state
$(S_i,\theta_i)$ is known at receiver $i$ but not known at the
transmitter. We will refer to the fading Gaussian BC defined by
\eqnref{FBC-complex} as the fading Gaussian BC
$(S_1,S_2)$.

Now suppose user $i$ channel has phase $\theta_i =\theta$. Since this
phase is known at the receiver, user $i$ can post-rotate
its received signal by $-\theta_i$.
We can write the complex signal input
as $\Xtil=(\re{X}+j\im{X})/\sqrt{2}$
and, following post-rotation of the phase,  we can represent  the
receiver $i$ additive noise as
$({\re{Z}}_i+j{\im{Z}}_i)/\sqrt{2}$, and the real and
complex components $\re{Y}$ and $\im{Y}$ of the receiver $i$ output
as
\begin{equation}
{\re{Y}}_i+j{\im{Y}}_i =\sqrt{2}e^{-j\theta_i}\Ytil_i= \sqrt{S_i} \re{X} +{\re{Z}}_i
+j(\sqrt{S_i}\im{X}+{\im{Z}}_i),\qquad i=1,2.
\end{equation}
Thus the in-phase and quadrature channels define a pair of
identical parallel fading broadcast channels, each with unit-power
additive Gaussian noise. We can assume without loss of
generality that signals $\re{X}$ and $\im{X}$ are independent and each
have unit power. Note this implies $\var{\Xtil}=1$. Henceforth, we will
evaluate a communication scheme on the real-valued broadcast channel
\begin{equation}\eqnlabel{FBC-real}
Y_i =  \sqrt{S_i} X + Z_i,\qquad i=1,2,
\end{equation}
that corresponds to the (post-rotated) in-phase and quadrature channels of
the complex fading Gaussian BC \eqnref{FBC-complex}.
The achievable rates and outer bounds of the complex fading Gaussian
BC  \eqnref{FBC-complex} will be precisely
double those obtained in the real-valued channel
\eqnref{FBC-real}.

\subsection{Fading Gaussian BC: Outer Bound}
Let $\C$ be the capacity region of the real-valued fading broadcast
channel \eqnref{FBC-real}.
Fix $\omega > 1$. We want to upper bound
\begin{equation}
R^*(\w)= \max_{(R_1,R_2) \in \C} R_1 + \omega R_2.
\end{equation}
As in the layered erasure BC, this channel is not degraded. However,
we can enhance user 1's channel to make it degraded just as we did for
the binary expansion channel.  The fading state of this enhanced user
is denoted $\Stil_1$ and has complementary CDF
\begin{equation}
\ccdf{\Stil_1}{s}=\limiter{\max(\ccdf{S_1}{s},\w\ccdf{S_2}{s})}.
\eqnlabel{S1enhanced}
\end{equation}
The resulting broadcast channel is now
degraded with user~2 as the weaker user. As before, we now can use
\Thmref{KM5} to write
\begin{align}\eqnlabel{Rstar-cts-start}
R^*(\w)&\le \max_{V,X} I(X;Y_1,\tilde{S}_1|V) + \omega I(V;Y_2,S_2).
\end{align}
To characterize $R^*(\w)$, we need a few definitions.  When channel
$i$ is in state $s$, receiver
$i$ observes an output identically distributed as
\begin{equation}
\Ys := \sqrt{s} X + Z,
\end{equation}
where $Z$ denotes a $N(0,1)$ random variable identical to each $Z_i$.
When the channel state  is a random process $S$, the
ergodic capacity of this point-to-point
Gaussian fading channel  with unit transmit power is
\begin{equation}
C_e(S):=\int_0^\infty \pdf{S}{s}\frac{1}{2}\log(1+s)\,ds
\eqnlabel{ErgodicCapacity}
\end{equation}
With $\ccdf{\Stil_1}{s}=\prob{\Stil_1\ge s}$ and
$\ccdf{S_2}{s}=\prob{S_2\ge s}$ denoting the complementary CDFs of
$\Stil_1$ and $S_2$ respectively,  we define
\begin{equation}
\ccdf{\w}{s}=\ccdf{\Stil_1}{s}-\w\ccdf{S_2}{s}.\eqnlabel{Fbar-defn}
\end{equation}
Finally, let $I'(X;\Ys|V)=dI(X;\Ys|V)/ds$ denote
the derivative of the conditional mutual information with respect to
the channel SNR $s$. Standard manipulations, as shown in the appendix,
yield the next claim.
\begin{lemma}\label{lem:Ideriv}
\begin{align}
R^*(\w)&\le \max_{V,X} \int_0^\infty
\ccdf{\w}{s}I'(X;\Ys|V)\,ds+\w C_e(S_2).
\end{align}
\end{lemma}
Lemma~\ref{lem:Ideriv} is the continuous-state version of
\Eqnref{RBsum} for the layered-erasure BC. In \eqnref{RBsum}, the
weighted sum rate is expressed in terms of the incremental
information from an improvement in channel state by one level. In
Lemma~\ref{lem:Ideriv}, $I'(X;\Ys|V)\,ds$ represents this same
incremental gain. Just as in layered erasure BC, we are now able to
optimize over the auxiliary $V$.

It was shown in \cite{GSV2005} that
\begin{equation}
I'(X;\Ys|V)=\frac{\log e}{2}\mmse{s|V},
\end{equation}
where, given a conditioning variable $V$,
\begin{equation}
\mmse{s|V}:=\E{(X-\E{X|\Ys,V})^2}
\end{equation}
denotes the mean square error of the conditional mean
estimator $\E{X|\Ys,V}$.
This implies
\begin{align}
R^*(\w)&\le \max_{V,X} \frac{\log e}{2}\int_0^\infty
\ccdf{\w}{s}\mmse{s|V}\,ds+ \w C_e(S_2).\eqnlabel{Rstar-mmse}
\end{align}
We note that $\mmse{s|V}\ge 0$ and that this minimum is achieved when
$V$ specifies $X$. In addition, we also have the upper bound
\begin{equation}
\mmse{s|V}\le \frac{1}{1+s}
\end{equation}
and this upper bound is achieved when $X\sim N(0,1)$, independent of
$V$. We thus obtain an upper bound to the right side of
\eqnref{Rstar-mmse} when
\begin{equation}
\mmse{s|V}=\begin{cases}
1/(1+s), & \ccdf{\w}{s}> 0,\\
0  & \ow.
\end{cases}
\end{equation}
Defining
\begin{equation}
\Itilset_1(\w)=\set{s\ge0|\ccdf{\w}{s}> 0}
=\set{s\ge 0| \ccdf{\Stil_1}{s}> \w\ccdf{S_2}{s}}
\end{equation}
and $\Itilset_2(\w)$ as its complement,  we can equivalently write
\begin{equation}
\mmse{s|V}=\begin{cases}
1/(1+s), & s\in\Itilset_1(\w),\\
0  & s\in\Itilset_2(\w).
\end{cases}\eqnlabel{mmse-bound}
\end{equation}
Applying \eqnref{Fbar-defn} and \eqnref{mmse-bound} to
\eqnref{Rstar-mmse}  yields
\begin{align}
  R^*(\w)&\le \frac{\log e}{2}\int_{\Itilset_1(\w)}
\ccdf{\w}{s}\frac{1}{1+s}\,ds+ \w C_e(S_2)\\
&=\frac{\log e}{2}\paren{\int_{\Itilset_1(\w)}
\ccdf{\Stil_1}{s}\frac{1}{1+s}\,ds
+ \w \int_{\Itilset_2(\w)} \ccdf{S_2}{s}\frac{1}{1+s}\,ds}.
\eqnlabel{Rstar-mmse2}
\end{align}
Defining
\begin{subequations}\eqnlabel{Iset-cts}
\begin{align}
\Iset_1(\w)&=\set{s\ge 0| \ccdf{S_1}{s}> \w\ccdf{S_2}{s}},\\
\Iset_2(\w)&=\set{s\ge 0| \ccdf{S_1}{s}\le \w\ccdf{S_2}{s}},
\end{align}
\end{subequations}
one can verify that
Lemma~\ref{lem:Isets} still holds for continuous fading
distributions; that is, $\Itilset_i(\w)=\Iset_i(\w)$. Moreover, if
$s\in\Itilset_1(\w)$, then $\ccdf{\Stil_1}{s}=\ccdf{S_1}{s}$.
It then follows from \eqnref{Rstar-mmse2} that
\begin{align}
R^*(\w)
&\le\frac{\log e}{2}\paren{\int_{\Iset_1(\w)}
\ccdf{S_1}{s}\frac{1}{1+s}\,ds
+ \w \int_{\Iset_2(\w)} \ccdf{S_2}{s}\frac{1}{1+s}\,ds}.
\eqnlabel{Rstar-mmse3}
\end{align}
Returning to the complex fading Gaussian BC in which rates are twice
those of the in-phase channel, we observe that the components of the
upper bound \eqnref{Rstar-mmse3} correspond to the extreme points of
the following outer bound.
\begin{theorem}\thmlabel{FBC-outer}
The capacity region of the fading Gaussian broadcast channel is
contained in the convex hull of the union over all $\w\ge0$ of rate
pairs $(R_1,R_2)$ satisfying
\begin{subequations}
\eqnlabel{R-outer}
\begin{align}
R_1&\le R_1^*(\w):=\log e\int_{\Iset_1(\w)}
\ccdf{S_1}{s}\frac{1}{1+s}\,ds,\\
R_2 &\le R_2^*(\w) :=\log e\int_{\Iset_2(\w)} \ccdf{S_2}{s}\frac{1}{1+s}\,ds.
\end{align}
\end{subequations}
\end{theorem}
We observe that for a point-to-point fading channel with unit transmit power
and receiver CSI $S$, the ergodic capacity \eqnref{ErgodicCapacity}
can be written as
\begin{equation}
C_e(S)=\log e\int_0^\infty \ccdf{S}{s}\frac{1}{1+s}\,ds.
\eqnlabel{ErgodicCapacity2}
\end{equation}
When $I_j(\w)$ is empty and all channel states
are ``assigned'' to receiver $i\neq j$, we see that the outer bound
for $R_i$ is
simply the ergodic capacity of the point-to-point fading channel to
receiver $i$. Thus  the outer bound rates \eqnref{R-outer}
are tight when the channel input is assigned to a single
receiver. However,  when there is a  partitioning by channel state of the
available ergodic capacity, we will see that the outer bound rates
\eqnref{R-outer} are loose in the absence of an achievability scheme
in which an auxiliary $V$ satisfies \eqnref{mmse-bound}.

\subsection{Fading Gaussian BC: Achievability}
\label{sect:DLGBC-achievability}
We will employ an achievable scheme based on
superposition of independent binary streams composed of $\pm1$
symbols. In particular, for  the
real-valued fading Gaussian BC \eqnref{FBC-real}, let the channel
input $X$ at time $t$ be given by 
\begin{equation}\eqnlabel{Xk-defn}
X[t] =\sqrt{3}\sum_{n=1}^\infty \Xtil_n [t] 2^{-n}
\end{equation}
where $\Xtil_1[t],\Xtil_2[t],\ldots$ are independent signals taking
values in $\set{-1,1}$ equiprobably. Each stream $\{\Xtil_n[t]\}$
will communicate an independent layer $n$ data stream, encoded at a
rate $r_n$ which is chosen to tolerate interference from the other
streams, channel variations and receiver noise. Note that unlike the
rest of the paper, we make explicit the dependency on the symbol
time $t$ to emphasize that the symbols within a layer are coded
across time. 

Each receiver will use a two-stage decoding procedure.
First, receiver $i$ employs a {\em detector} to form estimates
$\Xhat_1[t],\Xhat_2[t],\ldots$ over a single-symbol period of  the
antipodal binary symbols $\Xtil_1[t],\Xtil_2[t],\ldots$ at each time
instant $t$. In the second stage, receiver $i$ employs the estimates
$\Xhat_n[1],\Xhat_n[2],\ldots$ to decode the coded sequence
$\Xtil_n[1],\Xtil_n[2],\ldots$  for each layer $n$ stream that the
receiver is assigned.

As the channel state $S_i$ is known at 
receiver $i$, the receiver outputs can be normalized by the channel gains,
so that the real broadcast channel \eqnref{FBC-real} with input $X[t]$ given
by \eqnref{Xk-defn} is equivalent to 
\begin{equation}\eqnlabel{FBC-real2}
\Ytil_i[t]= \sum_{n=1}^\infty\Xtil_n[t]2^{-n} +
\frac{Z_i[t]}{\sqrt{3S_i}},\qquad i=1,2.
\end{equation}
In the broadcast channel, different layers are
assigned to different users. The code rate employed on a layer depends
on which receiver is assigned that layer, which in turn depends on the
channel state distributions $\ccdf{S_i}{s}$ and the
weight factor $\w$.
\begin{figure}\label{fig:binsup-tx}
\begin{center}
\includegraphics{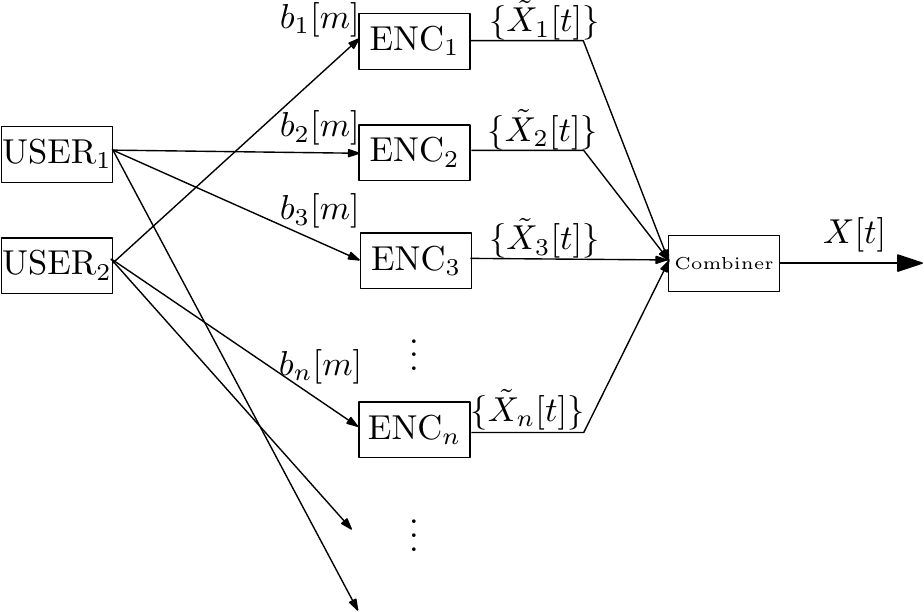}
\end{center}
\caption{In this example of the in-phase baseband transmitter, the
  layer $2$ and layer $3$ signals are assigned to user $1$ while the
  layer $1$ signal is assigned to user $2$. For signal layer $k$
  assigned to user $i$, the bit stream $b_k[n]$ is
  sent to the layer $k$ encoder at rate $r_{k,i}$. The transmitted
  signal $X[t]$ is given by \Eqnref{Xk-defn}.}
\end{figure}

 Before proceeding to the analysis of this system, we develop some
 terminology to describe our signaling scheme.  
We start by expressing the number $b\in[0,1]$ as the binary
expansion  $b=\sum_{n=1}^\infty b_n2^{-n}$ with each
$b_n\in\set{0,1}$. We denote such a binary expansion by
$b=\expansion[0,1]{b_1,b_2,\ldots}$, where the subscript $\set{0,1}$
denotes the range of each element.
Next we observe  $a=2b-1$ spans the interval
$[-1,1]$ and has the corresponding expansion
\begin{align}
a = 2b-1
&=2\sum_{n=1}^\infty b_n2^{-n} -\sum_{n=1}^\infty 2^{-n}
= \sum_{n=1}^\infty (2b_n-1)2^{-n}.
\end{align}
Defining $a_n=2b_n-1$, we see that  any $a\in[-1,1]$ can be expressed
in terms of the {\em antipodal expansion} 
$a=\sum_{n=1}^\infty a_n2^{-n}$
with each $a_n\in\set{-1,1}$. Moreover, given $a\in[-1,1]$, we can
construct the corresponding antipodal expansion, 
denoted $\expansion{a_1,a_2,\ldots}$,  
via $a_1=\sgn(a)$, $a_2=\sgn(a-a_1/2)$, and
\begin{equation}\eqnlabel{float2expansion}
a_{n+1}=\sgn\paren{a-\sum_{j=1}^n a_j 2^{-j}}.
\end{equation}
In certain situations, we will employ the signaling scheme
 \eqnref{Xk-defn}, but with only a finite number $m$ of signal
 levels. In this case,  the transmitted signal constellation for a
 single symbol period is
given by
\begin{equation}
\Xcal=\set{\sum_{n=1}^m x_n 2^{-n}|x_n\in\set{-1,1},n=1,2,\ldots,m}.
\end{equation}
That is, $\Xcal$ consists of the set of all $m$-bit antipodal
expansions $\expansion{x_1,\ldots,x_m}$.  In this case, a receiver may
observe $y\in[-1,1]$ and wish to determine the $\xhat\in\Xcal$ closest
to $y$. Expressing $y$ in terms of its antipodal expansion
$y=\sum_{n=1}^\infty y_n2^{-n}$, one can show that the closest
$\xhat\in \Xcal$ is $\xhat=\sum_{n=1}^m y_n2^{-n}$, which is simply
the truncation of the antipodal expansion of $y$ to its first $m$
bits. We refer to this as the {\em truncation property} of the
antipodal expansion.\footnote{We observe that the truncation property
  is not shared by the ordinary binary expansion. For example, $15/32$
  has the binary expansion $\expansion[0,1]{0,1,1,1,1,0,\ldots}$ but
  $\expansion[0,1]{1,0}=1/2$ and $\expansion[0,1]{1,0,0}$ are the
  $2$-bit and $3$-bit binary expansions closest to $15/32$. However,
  $15/32$ has the antipodal expansion $\expansion{1,-1,1,1,1}$ and
  $\expansion{1,-1}$ and $\expansion{1,-1,1}$ are the closest $2$-bit
  and $3$-bit antipodal expansions.}

Returning to the real-valued fading Gaussian BC \eqnref{FBC-real2},
consider a receiver for $\Ytil[t]$, which could be either $\Ytil_i[t]$
with channel state $S_i[t]=s$. To evaluate this receiver, we 
drop the time index $t$ and consider the first stage detection over a
single symbol period with channel state $S_i=s$. The received signal
is
\begin{equation}\eqnlabel{Ytil-defn}
\Ytil=\Xtil+\Ztil,
\end{equation}
where the transmitted signal constellation is the set of
all $\Xtil=\sum_{n=1}^\infty\Xtil_n2^{-n}\in [-1,1]$
and the receiver noise is $\Ztil=Z/\sqrt{3s}$. 
 For this receiver, the signal
set consists of all antipodal expansions 
$\Xtil=\expansion{\Xtil_1,\Xtil_2,\ldots}\in [-1,1]$. 
The receiver detects the input bits
$\{\Xtil_i\}$  by mapping the observation $\Ytil$ to the nearest signal
constellation point 
\begin{equation}
\Xhat=\expansion{\Xhat_1,\Xhat_2,\ldots}=\sum_{n=1}^\infty
\Xhat_n2^{-n}, \eqnlabel{Xhat-expansion}
\end{equation}
where
$\Xhat_n\in\set{\pm1}$ is the receiver's best estimate of the layer $n$
input bit $\Xtil_n$. This minimum distance detection reduces
to  
\begin{equation}\eqnlabel{Xhat-detector}
\Xhat=\max\paren{-1,\limiter{\Ytil}}.
\end{equation}
such that given $\Xhat$, we derive the antipodal bits $\{\Xhat_n\}$
using the method of \eqnref{float2expansion}.  That is, when
$\Ytil\in[-1,1]$, $\Xhat=\Ytil$ and the binary expansion of $\Ytil$ is
our best estimate for the bits $\{\Xtil_n\}$. When $\Ytil>1$, the
detector estimates $\Xhat_n=1$ for all $n$ and when $\Ytil<-1$, the
detector estimate $\Xhat_n=-1$ for all $n$.

As a notation convenience, we use $\Xtil^n$ to represent the
$n$-bit antipodal expansion $\expansion{\Xtil_1,\ldots,\Xtil_{n}}=\sum_{j=1}^n
\Xtil_j2^{-j}$.  This 
notation is also applied to $\Xhat^n$, $\Ytil^n$ and sample values
such as $\xtil^n$ of $\Xtil^n$.
For the detector \eqnref{Xhat-detector}, 
we observe that the probability $\prob{\Xhat_n=\Xtil_n}$
of a correct bit decision at layer $n$ is lower bounded by 
\begin{equation}\eqnlabel{pc-lowerbound}
\prob{\Xhat_n=\Xtil_n}\ge \prob{\Xhat^n=\Xtil^n}. 
\end{equation}
To proceed, we will find a lower bound to
$\prob{\Xhat^n=\xtil^n|\Xtil^{n}=\xtil^n}$ that holds for all $\xtil^n$.
Given $\Xtil^n=\xtil^n$, we can write
$\Xtil=\xtil^n + U_n,$ where
\begin{equation}
U_n=\sum_{j={n+1}}^\infty \Xtil_j2^{-j}
\end{equation}
is  a continuous uniform $(-2^{-n},2^{-n})$ random variable that is
independent of $\Xtil^n$ and $\Ztil$.  We refer to $U_n$ as {\em LSB
interference} since at level $n$ it is the superposition of signals
for the less significant bits $\{\Xtil_j|j>n\}$. 

Thus, given $\Xtil^n=\xtil^n$, the receiver observes
\begin{equation}\eqnlabel{Ytil-conditional}
\Ytil=\xtil^n+U_n+\Ztil.
\end{equation}
From \eqnref{Xhat-detector} and the truncation property of the
  antipodal expansion, $|\Ytil-\xtil^n|< 2^{-n}$ implies 
$\Xhat^n=\xtil^n$.  With the definition of the conditional
probability 
\begin{align}\eqnlabel{pcn2}
\prob{\xtil^n-2^{-n} \le
\Ytil<\xtil^n+2^{-n}|\Xtil^{n}=\xtil^n},
\end{align}
it follows that 
\begin{equation}
\prob{\Xhat^n=\xtil^n|\Xtil^{n}=\xtil^n}\ge \pc{\xtil^n}.
\end{equation}
From \eqnref{Ytil-conditional}, \eqnref{pcn2} and independence of
$\Xtil^n$, $U_n$, and $\Ztil$, 
\begin{align}
\pc{\xtil^n} &=\prob{-2^{-n}\le U_n+\Ztil< 2^{-n}},
\eqnlabel{noncrossprob}
\end{align}
which is independent of $\xtil^n$. 

When the fading state is reasonable, the lower bound $\pc{\xtil^n}$ is
sufficient to characterize the reliability of the bit detection at
layer $n$.  However, as the fading state becomes too weak, this lower
bound will approach zero, corresponding to our detector converging to
random guessing.  Analysis of the detector for weak fading states is
complex but, fortunately, unnecessary. As the channel state $S=s$ is
known at the receiver, we can simplify our analysis by assuming the
detector sets a threshold $\nhat(s)$ such that for layers $n>\nhat(s)$
the receiver simply sets $\Xhat_n$ to a random guess.
To be precise,  a receiver with observation
$\Ytil$ in channel state $S=s$ implements the first stage detector
\begin{subequations}\eqnlabel{binary-detector}
\begin{align}
\expansion{\Yhat_1,\Yhat_2,\ldots}&=\max\paren{-1,\limiter{\Ytil}},\\
\Xhat_n&=\begin{cases}
\Yhat_n & 1\le n\le \nhat(s),\\
W_n & n> \nhat(s),
\end{cases}
\end{align}
\end{subequations}
where $W_1,W_{2},\ldots$ is simply an iid sequence of
equiprobable Bernoulli random variables, independent of any
observations. We choose the threshold $\nhat(s)$ as the largest $n$
such that $\pc{\xtil^n}\ge 1/2$. This provides a simple guarantee 
that our detector never does worse than random guessing.

The exact calculation of $\pc{\xtil^n}$ is
shown in a generalized form in the proof of
Lemma~\ref{lem:crossover-d}.  We  will express the result in terms of
\begin{align}
a_{n}(s)&:=
3s2^{-2n},\\
G(x) &:= xQ(x) -\frac{1}{\sqrt{2\pi}}e^{-x^2/2},\\
\epsilon_d(a)&:=\frac{G(\sqrt{a}(1+2^{-d}))-G(\sqrt{a}(1-2^{-d}))}{\sqrt{a}2^{-d}}
\eqnlabel{ed-defn}
\intertext{and}
\epshat_d(a)&:=\limiter[1/2]{\epsilon_d(a)}.
\end{align}
We note that $a_n(s)$ is the SNR of bit $\Xtil_n$ under channel
state $s$ and that $G(x)$ is simply the integral of $Q(x)$.
In the proof of Lemma~\ref{lem:crossover-d}, we show
$1-\pc{\xtil^n}=\epsilon_0(a_n(s))$. By using random guessing for bits
$\{\Xtil_n|n>\nhat(s)\}$, the next lemma follows. 
\begin{lemma}\label{lem:crossover}
In channel state $S=s$, the bit detector \eqnref{binary-detector}
yields a BSC from $\Xtil_n$ to $\Xhat_n$ with crossover probability
$\pcross_{n,0}(s) \le\epshat_{0}(a_n(s))$.
\end{lemma}
This is  the $d=0$  special case of Lemma~\ref{lem:crossover-d}
in Section~\ref{sect:gap-reduction}.
It is straightforward to verify that
$\epsilon_0(a)=[G(2\sqrt{a})-G(0)]/\sqrt{a}$ is
decreasing in $a$ and that $\epsilon_0(a=0.5405)=1/2$.
The upper bound of $1/2$ inherent in the definition of $\epshat_0(a)$
comes from defining the threshold
\begin{equation}\eqnlabel{nhat-threshold}
\nhat(s)=\max\set{n|\epsilon_{0}(a_n(s)) \le
  1/2}\approx\floor{0.349+\frac{1}{2}\log s}.
\end{equation}
The threshold $\nhat(s)$ is chosen simply so that for bits at
level $n\le \nhat(s)$, the effective channel from $\Xtil_n$ to the
receiver guess $\Xhat_n$ is a binary symmetric channel (BSC) with
crossover probability less than $1/2$.

The implication of Lemma~\ref{lem:crossover} is that receiver $i$
observes level $n$ bits through a BSC with time-varying crossover
probability $\pcross_{n,0}(S_i)$. We employ this signaling and
detection scheme  on both the in-phase and quadrature channels,
coupled with coding over time. With $H(p)$ denoting the binary entropy
function, user $i$ can communicate reliably using the level $n$
channel at rate
\begin{align}\eqnlabel{raten-1}
\rate{n}&\ge 2\E[S_i]{1-H(\pcross_{n,0}(S_i))}
\ge 2\E[S_i]{1-H(\epshat_0(a_n(S_i)))}.
\end{align}
For convenience, we define
\begin{equation}
\entropyhat{s}:=1-H(\epshat_{d}(s)).
\end{equation}
In our achievability scheme, we are free to assign each signal level $n$
arbitrarily. By considering the assignments of signal levels
$n\in\hat{\Nset}_i$
to each receiver, we can achieve the following rate region.
\begin{theorem}\thmlabel{FBC-inner-bound}
 The capacity region of the fading
  Gaussian BC $(S_1,S_2)$ includes all rate pairs
$(R_1,R_2)$ satisfying
\begin{align*}
R_i&\le  2\sum_{n\in\hat{\Nset}_i}
\E[S_i]{\entropyhat[0]{3S_i2^{-2n}}},
\qquad i=1,2,
\end{align*}
for some partition
$\hat{\Nset}_1\cup\hat{\Nset}_2$ of  $\set{1,2,\ldots}$.
\end{theorem}
In Section~\ref{sect:const-gap}, we show that the achievable rates of
\Thmref{FBC-inner-bound} are within a constant gap of the
\Thmref{FBC-outer} outer bound. However, \Thmref{FBC-inner-bound} does
not exploit knowledge of the channel fading distributions. In
the following subsection, we describe how a receiver can offer a
considerable improvement in rates for specific channel state distributions.
\subsection{Improved Rates via Reverse Stripping}
\label{sect:gap-reduction}
While the binary expansion superposition achievability scheme in
Section~\ref{sect:DLGBC-achievability} mimics the
structure of the layered erasure channel, it has two
disadvantages:
\begin{itemize}
\item At level $n$, the detector makes a hard decision on the bits at
  levels $1$   through $n$. These decisions do not exploit coding
  over time.

\item The hard decisions at level $n$ are subject to {\em LSB
  interference} from bits at levels $k>n$.
\end{itemize}
A consequence of the detector's hard decisions is that binary
expansion superposition sacrifices low SNR performance.
In particular, \eqnref{nhat-threshold} shows that the level $n$ bit is
useless for channels state $s <2^{2(n-0.349)}$, implying no channel
state $s\le 2^{1.302}$ communicates information.
While we are perhaps stuck with the hard decisions, the LSB interference
can be mitigated in some circumstances.  To explain this, we suppose
initially that the weight $\omega$ is chosen so that all layers are
assigned to user $i$. Although user $i$ could obtain his
point-to-point ergodic capacity rate via constant power Gaussian
codes, it's instructive
to examine what can be achieved with the binary expansion
superposition scheme. In this case, we could do the following {\em reverse stripping}:

\begin{itemize}
\item Select a level $n_{\max}$, corresponding to the least significant
  bit we will transmit, so that the forfeited capacity
is negligible.

\item At layer $n=\nmax$, make hard decisions $\Xtil_{n}[t]$ using the BES
 detector \eqnref{binary-detector}. In this case, the
  layer $\nmax$ crossover probability is reduced due to the absence
  of LSB interference.

\item After a code block has been sent, decode the level $n$ codeword and
then strip the level $n$ signal from the received signal.

\item Go back and redo the binary detection for level $n-1$ using the
residual received signal. Now detection at level $n-1$ is no longer
subject to LSB interference from level $n$.

\item Repeat this process all the way down to bit $X_1$.
\end{itemize}
We call this process {\em reverse stripping} because we decode and
strip the signals in the order of increasing signal power, which
is the reverse of the decoding order for Gaussian superposition codes.

At each level $n$, this process reduces the crossover probability for
the binary detector \eqnref{binary-detector} for a given channel state
$S_i=s$. In particular, in the
error probability analysis of equations~\eqnref{Ytil-defn}
through~\eqnref{noncrossprob}, the received signal is still
$\Ytil=\Xtil+\Ztil$, but the transmitted signal is now a constellation point
\begin{equation}
\Xtil=\expansion{\Xtil_1,\ldots,\Xtil_{\nmax}}
=\sum_{j=1}^{\nmax}\Xtil_j2^{-j}.
\end{equation}
For detection of bit $n=\nmax$, \eqnref{pc-lowerbound} and
\eqnref{pcn2} remain unchanged; however, given $\Xtil^n=\xtil^n$, we
now have $\Ytil=\xtil^n +\Ztil$. That is, the LSB interference $U_n$
is now zero. Following \eqnref{noncrossprob}, the probability of
correct detection of the bit $X_n$ for  $n=\nmax$ is lower bounded by
\begin{align}
\pc{\xtil^n} &=\prob{-2^{-n}\le \Ztil< 2^{-n}}
=1 - 2 Q(\sqrt{a_n(s)}).\eqnlabel{noncrossprob2}
\end{align}
We note from \eqnref{ed-defn} that
\begin{equation}\eqnlabel{Gderiv}
\epsilon_{\infty}(a):=\limty{d}\epsilon_d(a)
= 2\eval{\derivf{}{G}{x}}_{x=\sqrt{a}}=2Q(\sqrt{a}).
\end{equation}
In addition, we define
$\epshat_{\infty}(a)=\limiter[1/2]{\epsilon_{\infty}(a)}$.
Denoting the crossover probability at level
$n$  by $\pcross_{n,\infty}(s)$, \eqnref{noncrossprob2} and
\eqnref{Gderiv} yield the upper bound
\begin{equation}\eqnlabel{pcn-infty}
\pcross_{n,\infty}(s)\le 1-\pc{\xtil^n}=\epshat_{\infty}(a_n(s)).
\end{equation}

Following the reverse stripping strategy, we strip the coded signal
$\Xtil_{\nmax}[t]$. The residual transmitted signal is now
$\Xtil=\sum_{n=1}^{\nmax-1} \Xtil_n2^{-n}$. Repeating the same
analysis, the error probability for bit $n=\nmax-1$ also satisfies the
upper bound \eqnref{pcn-infty}. By successive reverse stripping, the
upper bound \eqnref{pcn-infty} holds for all bits $\Xtil_n$.
It follows that user $i$ achieves the ergodic rate
\begin{align*}R_i=  2\sum_{n=1}^\infty
\E[S_i]{\entropyhat[\infty]{3S_i2^{-2n}}}.
\end{align*}


Reverse stripping also can be exploited for binary expansion
superposition signaling in the general setting when the bits are
assigned to different users. Here suppose that user $i$ is  assigned
bit levels
\begin{equation}\eqnlabel{Nset-d}
\Nset_i=\bigcup_{j=1}^k\set{n_j,n_j+1,\ldots, l_j},
\end{equation}
where $n_j\le l_j< n_{j+1}$.
In this case, we can apply reverse stripping to decoding each interval of bits
$\{\Xtil_{n_j},\ldots, \Xtil_{l_j}\}$. Starting with bit $n=l_j$,
the probability of correct decoding of bit $\Xtil_{n}$ is lower
bounded by
\begin{align}\eqnlabel{pcnd1}
\pc{\xtil^n} &= \prob{\Ytil^n=\xtil^n|\Xtil^{n}=\xtil^n}.
\end{align}
Given $\Xtil^n=\xtil^n$, we can write  $\Xtil=\xtil^n+U_n$ with
\begin{equation}\eqnlabel{LSBinterference}
U_n=\sum_{j={n+1}}^\infty \Xtil_j2^{-j}
\end{equation}
denoting the LSB interference. In fact, user $i$ will have previously
decoded (via reverse stripping) the LSBs $\cup_{k=j+1}^\infty
\{\Xtil_{n_k},\ldots,\Xtil_{l_k}\}$ and these bits can be stripped
from the LSB interference $U_n$. However, $U_n$ also contains bits
$\set{X_{n+1}\,\ldots, X_{n_{j+1}-1}}$ that are assigned to the other
user and these bits may be encoded at a rate that user $i$ cannot
decode and strip reliably. Consequently, we assume that these bits are
undecodable by user $i$. Moreoever, as these same bits that are the
most significant bits in the LSB interference, they dominate the LSB
interference. Hence, there is only a small penalty in the assumption
that the LSB interference $U_n$ in \eqnref{LSBinterference} contains
no known bits and is thus statistically identical to a continuous
uniform $(-2^{-n},2^{-n})$ random variable. To be precise, we obtain a
lower bound to the probability of correct decoding because the LSB
interference $U_n$ with some known bits $\Xtil_j$ can be degraded by
replacing those known (and stripped) bits with antipodal $\pm1$ noise
to create the continuous uniform $U_n$. With this assumption, analysis
of the detection of bit $n=l_j$ is identical to that described in
equations \eqnref{pc-lowerbound} through \eqnref{noncrossprob} and the
crossover probability is given by Lemma~\ref{lem:crossover}.

Now suppose in the reverse stripping process we are at level $n=l_j-d$
and we wish to detect bit $\Xtil_n=\Xtil_{l_j-d}$ having already
decoded and stripped bits $\Xtil_{l_j-d+1}$ through $\Xtil_{l_j}$.
 In this case, given $\Xtil^n=\xtil^n$, the
residual received signal is
\begin{equation}\eqnlabel{Ytil-conditional-d}
\Ytil=\xtil^n+U_n+\Ztil.
\end{equation}
The residual LSB interference, after stripping the LSB interference at
levels $l_j$ through $l_{j-d+1}$, is
\begin{equation}
U_n=\sum_{k=l_j+1}^\infty\Xtil_k2^{-k}
=\sum_{k=n+d+1}^\infty \Xtil_k 2^{-k}
=2^{-d}\Util_n
\end{equation}
where
\begin{equation}
\Util_n=\sum_{k'=n+1}^\infty \Xtil_{k'+d}2^{-k'}.
\end{equation}
With respect
to the detection of bit $\Xtil_n$ for user $i$, the most significant
bits in $\Util_n$ are assigned to the other user and thus cannot be
assumed to be decodable by user $i$. Hence we can lower bound the
probability of correct detection by making
the worst-case assumption that $\Util_n$ is a continuous uniform
$(-2^{-n},2^{-n})$ random variable. Nevertheless, as $\Util_n$ is
identical to the LSB interference at layer $n$ without reverse
stripping, we say that the LSB interference
$U_n=2^{-d}\Util_n$ is at depth $d$. In this case, the probability of
correct detection at level $n$ with LSB interference at depth $d$ is
lower bounded by
\begin{align}\eqnlabel{pcnd2}
\pc{\xtil^n} &= \prob{\xtil^n-2^{-n}\le
  \Ytil<\xtil^n+2^{-n}|\Xtil^{n}=\xtil^n}\\
&=\prob{-2^{-n}\le 2^{-d}\Util_n+\Ztil<2^{-n}}
\eqnlabel{noncrossprob-d}
\end{align}
In the proof of the following Lemma, we show that
$1-\pc{\xtil^n}=\epsilon_d(a_n(s))$. 
In channel state $s$, we guess bits $\Xhat_n$ for layers
$n>\nhat(s)$, yielding an upper bound of $\epshat_d(a_n(s))$ for the
crossover probability at all layers $n$.  

\begin{figure}
\begin{center}\setlength{\unitlength}{0.1in}
\begin{picture}(50,35)
\put(5,2){\makebox(0,0)[bl]{\includegraphics{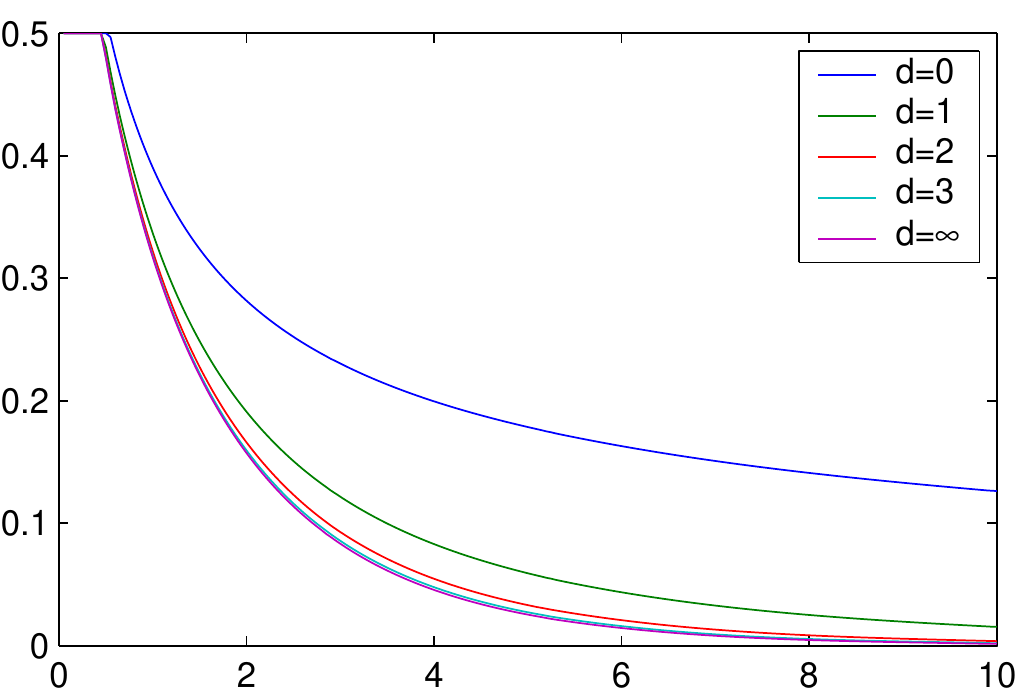}}}
\put(0,20){$\epsilon_d(a)$}
\put(30,0){$a$}
\end{picture}
\end{center}
\caption{In channel state $s$, the probability of bit error for the
  BES detector at level $n$ with LSB interference at depth $d$ is
  $\epshat_d(a)$ for $a=a_n(s)=3s2^{-2n}$.} 
\label{fig:edhat}
\end{figure}

\begin{lemma}\label{lem:crossover-d}
In channel state $S=s$ with LSB interference at depth $d$, the bit
 detector \eqnref{binary-detector}  yields a BSC from $\Xtil_n$ to
 $\Xhat_n$ with crossover probability
$\pcross_{n,d}(s) \le\epshat_{d}(a_n(s))$.
\end{lemma}

We note that the crossover probability of Lemma~\ref{lem:crossover} is
the special case of LSB interference at depth $d=0$. Similarly, when
all bits are assigned to user $i$ and there is no LSB interference,
the crossover probability $\epshat_{\infty}(a_n(s))$ given in
\eqnref{pcn-infty} corresponds to LSB interference at depth $d\goes\infty$.
In Figure~\ref{fig:edhat}, we plot $\epshat_d(a)$ and
$\epshat_\infty(a)$. We see there is a considerable improvement in
detection when we go from LSB interference at depth $d=0$ to
$d=1$ and that LSB interference at depth $d\ge3$ is essentially
indistinguishable from no LSB interference.

We now apply Lemma~\ref{lem:crossover-d}. Given a weight factor $\w$,
we assume that bits $n\in \Nset_i$ given by \eqnref{Nset-d} are
assigned to user $i$. At each level $n$, the interference must
overcome LSB interference at a depth $d_n$ that is specified by the
assignments $\Nset_i$. In particular, if $n\in N_i$, then
\begin{equation}
d_n=\min\set{d\ge0|n+d+1\not\in N_i}.
\end{equation}
With reverse stripping, receiver $i$
observes the bits of layer $n\in N_i$ through a BSC with time-varying
crossover probability $\pcross_{n,d_n}(S_i)$, enabling
reliable communication at rate
\begin{align}\eqnlabel{raten-d}
\rate{n}&\ge 2\E[S_i]{1-H(\pcross_{n,d_n}(S_i))}.
\end{align}
By partitioning the signal levels, we obtain the following achievable
rate region.
\begin{theorem}\thmlabel{FBC-inner-bound-rs}
 The capacity region of the fading
  Gaussian BC $(S_1,S_2)$ includes all rate pairs
$(R_1,R_2)$ satisfying
\begin{align}R_i&\le  2\sum_{n\in\hat{\Nset}_i}
\E[S_i]{\entropyhat[d_n]{3S_i2^{-2n}}},
\qquad i=1,2,
\end{align}
for some partition
$\hat{\Nset}_1\cup\hat{\Nset}_2$ of  $\set{1,2,\ldots}$.
\end{theorem}
\subsection{Examples}\label{sect:awgn-examples}
In this section, we compare achievable rates and the outer bound of
\Thmref{FBC-outer} for some simple examples. Beyond an initial
degraded AWGN BC example, the subsequent examples consider
non-degraded broadcast channels which, by construction, exhibit
signficant gains over time sharing. Our approach is to evaluate the
outer bound of
 \Thmref{FBC-outer}  and use the associated partition of signal
 levels to guide the assignment of BES bit levels in the achievability
 scheme.
\subsubsection*{Intermittent AWGN channels}
First we consider channels in
which each user $i$ has channel state $S_i$ described by
\begin{equation}\eqnlabel{ccdf-intermittent}
\ccdf{S_i}{s}=\begin{cases}
1 & s<0,\\
p_i & 0\le s \le s_i^*,\\
0 & s>s_i^*.
\end{cases}
\end{equation}
That is, each user $i$ has an {\em intermittent AWGN channel} in which
the SNR is $s_i^*$ with probability $p_i$ and is
otherwise zero. We
refer to $p_i$ as the channel activity factor and $s_i^*$ as the
maximum SNR. In all such
examples, we assume without loss of generality that $s_2^*\le s_1^*$.
We make the further assumption that $p_2 \ge p_1$. That is, the
intermittent good
channel of user $2$ is less good than that of user $1$ but user
$2$ more often has a good channel. It will be convenient
to express our results in terms of the ratio $\rho=p_1/p_2$ and the
ergodic capacity
\begin{equation}
C_i=p_i\log(1+s_1^*)
\end{equation}
that user $i$ would obtain with the full devotion of the transmitter's
resources.

For the outer bound, the partition
\eqnref{Iset-cts} yields
\begin{subequations}
\begin{align}
\Iset_1(\w)&=\begin{cases}
[0,s_1^*] & 0 \le \w <\rho,\\
(s_2^*,s_1^*] & \rho \le \w,
\end{cases}\\
\Iset_2(\w)&=\begin{cases}
(s_1^*,\infty) & 0 \le \w <\rho,\\ {}
[0,s_2^*]\cup(s_1^*,\infty) & \rho\le \w.
\end{cases}
\end{align}
\end{subequations}
It follows from \Thmref{FBC-outer} that the outer bound region has
extreme points
\begin{subequations}\eqnlabel{awgn-outer}
\begin{align}
R_1^*(\w) &= \begin{cases}
C_1 & 0 \le \w<\rho,\\
C_1-\rho C_2 & \rho \le \w,
\end{cases}\\
R_2^*(\w) &=\begin{cases}
0 & 0 \le \w<1,\\
C_2 & 1\le \w.
\end{cases}
\end{align}
\end{subequations}
Note that \eqnref{awgn-outer} identifies the extreme points $(C_1,0)$
and $(C_1-\rho C_2,C_2)$. Note that the complete outer bound region
also includes the corner point $(C_1,0)$ that is otherwise dominated
in the weighted sum rate $R_1+\w R_2$ by the extreme point $(C_1-\rho
C_2,C_2)$.

\begin{figure}[t]
\setlength{\unitlength}{0.06 in}
\begin{center}
\begin{tabular}{cc}
\Rplot{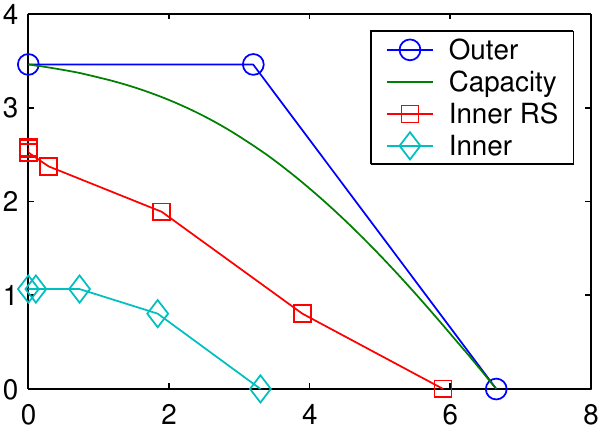}
 & \Rplot{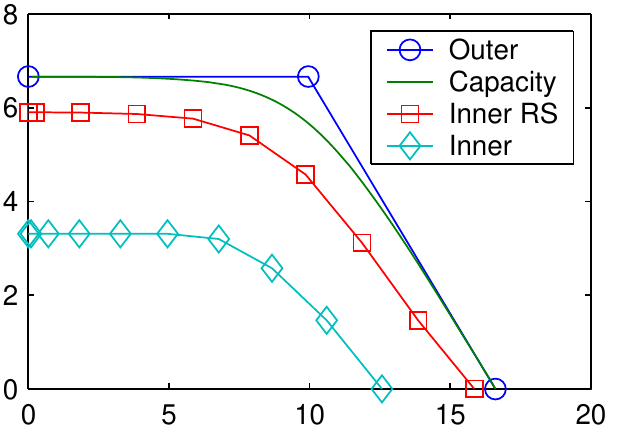}\\
{\bf (a)} & {\bf (b)}
\end{tabular}
\end{center}
\caption{AWGN broadcast channel rate regions: (a) users $1$ and $2$
  have SNRs $20$~dB and $10$~dB. (b) users $1$ and $2$ have SNRs
  $50$~dB and $20$~dB. Each plot compares the outer bound, the
  capacity, and the inner bounds with reverse stripping (RS) and without.}
\label{fig:awgn-bc}
\end{figure}

For the achievable rates of BES signaling, we observe from
\eqnref{nhat-threshold} that signals levels $1$ through
$n_2^*\approx(1/2)\log s_2^*$ are useful to both receivers while
levels above $n_2^*$ can transmit data only to receiver $1$. However,
we note that using those high levels for user $1$ does subject user
$2$ to the penalty of additional LSB interference. In particular, to
maximize the gain from reverse stripping, we assign levels up to
$n_2$ to user $2$ and levels above $n_2$ to user $1$. Thus user $1$
faces zero LSB interference. By varying
$n_2$, we obtain from \Thmref{FBC-inner-bound-rs} the boundary points
$(R_1,R_2)$ of an achievable rate region  given by
\begin{subequations}\eqnlabel{awgn-inner-rs}
\begin{align}
R_1 &= 2p_1\sum_{n=n_2+1}^\infty
\entropyhat[\infty]{3s_1^*2^{-2n}}\\
R_2 &= 2p_2\sum_{n=1}^{n_2}\entropyhat[n_2-n]{3s_2^*2^{-2n}}
\end{align}
\end{subequations}
Note that if reverse stripping is not implemented at the receivers,
the achievable rates are reduced to
\begin{subequations}\eqnlabel{awgn-inner-nrs}
\begin{align}
R_1 &= 2p_1\sum_{n=n_2+1}^\infty
\entropyhat[0]{3s_1^*2^{-2n}}\\
R_2 &= 2p_2\sum_{n=1}^{n_2}\entropyhat[0]{3s_2^*2^{-2n}}
\end{align}
\end{subequations}
For numerical comparisons,  we start with the ordinary AWGN
BC in which $p_1=p_2=1$ in order to assess the inner and outer bounds
when the capacity region is known \cite{Ber74}. Figure~\ref{fig:awgn-bc} considers high SNR examples in which (a) user 1
has SNR $10\log_{10} s_1^*=20$~dB while user $2$ has SNR $10$~dB and
(b) user $1$ has SNR $50$~dB and user $2$ has SNR $20$~dB.
Each plot compares the outer bound \eqnref{awgn-outer}, the capacity
given by Gaussian superposition codes, the inner bound
\eqnref{awgn-inner-rs} with reverse stripping and the inner bound
\eqnref{awgn-inner-nrs} without reverse  stripping. We see that the
outer bound is within one bit of the capacity boundary, an
observation appearing first in \cite{ADT-deterministic-allerton}. With
reverse stripping, the BES achievable rates are also within one bit of
the capacity region. However, in the absence of reverse stripping, the
 BES scheme is penalized considerably, although this penalty is
 dimishing with increasing SNR.

\begin{figure}[t]
\begin{center}\setlength{\unitlength}{0.06in}
\begin{tabular}{cc}
\Rplot{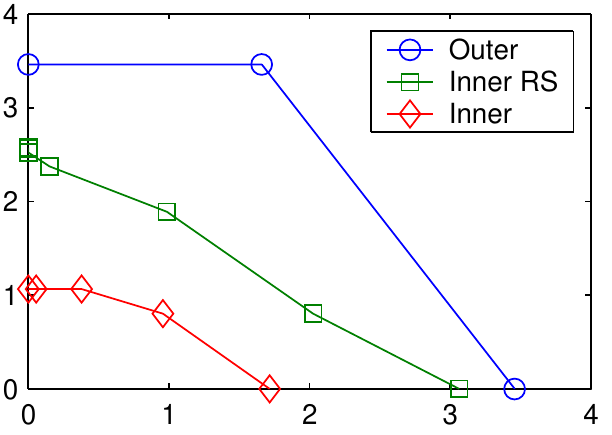} & \Rplot{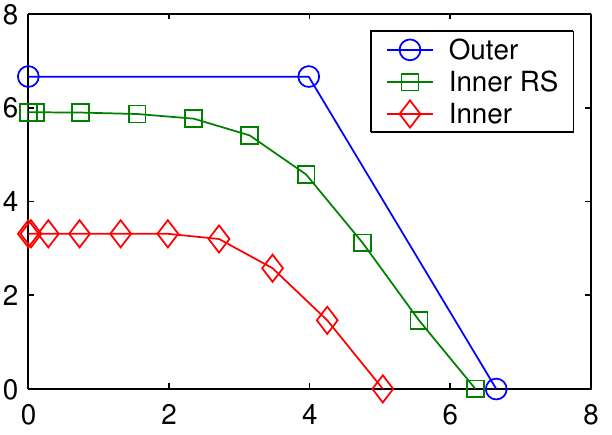}\\
{\bf (a)} & {\bf (b)}
\end{tabular}
\end{center}
\caption{AWGN broadcast channel rate regions: (a) users $1$ and $2$
  have SNRs $20$~dB and $10$~dB. (b) users $1$ and $2$ have SNRs
  $50$~dB and $20$~dB. In each case, the user $2$ channel is good with
  probability $p_2$. Each plot compares the outer bound, the
  capacity, and the inner bounds with reverse stripping (RS) and without.}
\label{fig:awgn-p}
\end{figure}

Next we consider a corresponding pair of broadcast channels in which
user $2$ has the same AWGN channel with $p_2=1$ but user $1$ now has
a channel activity  factor $p_1<1$ such that users $1$ and $2$ have
equal ergodic capacities $C_1=C_2$. In this case, numerical
comparisons of the inner and outer bounds are given in
Figure~\ref{fig:awgn-p}. In these cases, user $1$ has activity
factor (a) $p_1\approx 0.52$    and (b) $p_1=0.4$,
In comparing Figures~\ref{fig:awgn-bc}
and~\ref{fig:awgn-p}, we see little qualititative
difference. Essentially, both the inner and outer bounds reflect that
user $2$ has access to a high SNR channel with probability $p_1$. This
activity factor simply scales the bits rates achieved by user $2$. The
fundamental policy that the LSBs are reserved for user $2$ remains the
same.

\subsubsection*{Intermittent AWGN channel vs.~Rayleigh fading channel}
Now consider the case when user $1$ still has an intermittent AWGN
channel described by the activity probability $p_1$ and good channel SNR
$s_1^*$ but user $2$ has a Rayleigh fading channel state $S_2$ with
average SNR $\Gamma_2$ and thus complementary CDF
\begin{equation}\eqnlabel{ccdf-Rayleigh}
\ccdf{S_2}{s} =\begin{cases}
1 & s <0,\\
e^{-s/\Gamma_2} & s\ge 0.
\end{cases}
\end{equation}

\begin{figure}[t]
\begin{center}\setlength{\unitlength}{0.1in}
\Rplot{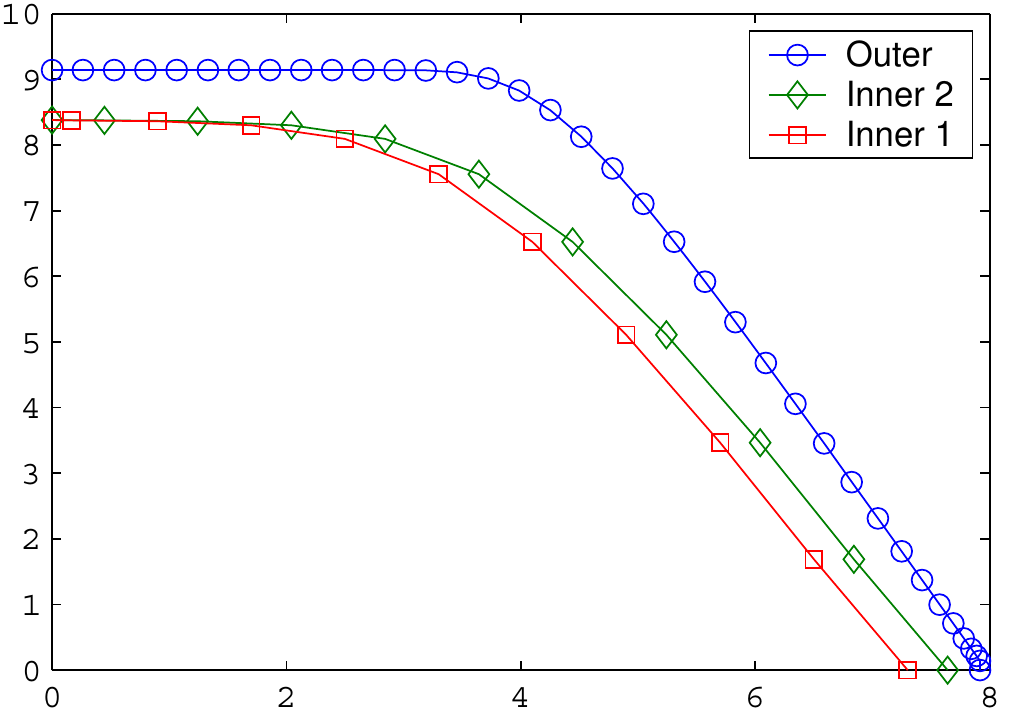}
\end{center}
\caption{Inner and outer bounds when user $1$ has an intermittent AWGN
  channel with activity factor $0.4$ and maximum SNR $60$~dB and user
  $2$ has a Rayleigh fading channel with $30$~dB average SNR.}
\label{fig:awgnray}
\end{figure}

For the outer bound, the partition
\eqnref{Iset-cts} yields
\begin{subequations}\eqnlabel{Iset-rayawgn}
\begin{align}
\Iset_1(\w)&=\begin{cases}
[0,s_1^*] & 0 \le \w <p_1,\\
(\Gamma_2\ln(\w/p_1),s_1^*] & p_1\le \w<p_1e^{s_1^*/\Gamma_2},\\
\phi & e^{s_1^*/\Gamma_2} \le \w,
\end{cases}\\
\Iset_2(\w)&=\begin{cases}
(s_1^*,\infty) & \w <p_1\\ {}
[0,\Gamma_2\ln(\w/p_1)]\cup(s_1^*,\infty) & p_1\le\w <
p_1e^{s_1^*/\Gamma_2},\\ {}
[0,\infty) & p_1e^{s_1^*/\Gamma_2}\le\w.
\end{cases}
\end{align}
\end{subequations}
We see in \eqnref{Iset-rayawgn} that varying  the weight
$\w$ over $[p_1,p_1e^{s_1^*/\Gamma_2}]$ simply corresponds to varying
a threshold channel state
\begin{equation}
s_\w := \Gamma_2\ln\frac{\w}{p_1}
\end{equation}
over the interval $[0,s_1^*]$. This permits us to write the channel
state partition as
\begin{subequations}\eqnlabel{Iset-rayawgn-s}
\begin{align}
\Iset_1(s_\w)&=(s_\w,s_1^*] \\
\Iset_2(s_\w) &=[0,s_\w]\cup(s_1^*,\infty)
\end{align}
\end{subequations}
To describe the outer bound region of \Thmref{FBC-outer}, we
define the integral function
\begin{equation}
\mu_{\Gamma}(a,b):=\log e\int_a^b e^{-s/\Gamma}\frac{1}{1+s}\,ds.
\end{equation}
The outer bound region is
then specified by a continuous boundary  of
extreme points $(R_1^*(s_\w),R_2^*(s_\w))$ found by varying $s_\w$
over the interval  $[0,s_1^*]$. From
\eqnref{Iset-rayawgn}, these boundary points are given by
\begin{subequations}\eqnlabel{awgn-rayleigh-outer}
\begin{align}
R_1^*(s_\w)
&= p_1\log\bracket{\frac{1+s_1^*}{1+s_\w}},\\
R_2^*(s_\w)
&=\mu_{\Gamma_2}(0,s_\w) +\mu_{\Gamma_2}(s_1^*,\infty).
\end{align}
\end{subequations}
This outer bound is shown in Figure~\ref{fig:awgnray} for an instance
in which receiver 1 has an intermittent AWGN channel $S_1$
with activity probability $p_1=0.4$ and
maximum SNR $s_1^*$ of $60$~dB.

For the BES achievability rates, we observe that user $1$ has maximum
SNR $s_1^*$ and thus only signal levels $n\le n_1^*:=\nhat(s_1^*)$ can
be used to communicate to receiver $1$. Hence signal levels $n>n_1^*$
are assigned to user $2$. However, signal levels $n\le n_1^*$ may be
useful for communication to either receiver. In particular, we observe
from the outer bound that when the weight $\w$ is sufficiently large,
signal levels $s$ below a threshold were associated with receiver $2$.
Hence for our achievability scheme, we assign signal levels
$n\in\set{1,\ldots,n_2}$ to user $2$ and signal levels
$n\in\set{n_2+1,\ldots,n_1^*}$ to user $1$. That is, for a given
threshold $n_1$, we obtain achievable rates from
\Thmref{FBC-inner-bound-rs} with
\begin{subequations}
\begin{align}
\hat{\Nset}_1&=\set{n_2+1,\ldots,n_1^*},\\
\hat{\Nset}_2&=\set{1,\ldots,n_2}\cup\set{n_1^*+1,n_1^*+2,\ldots}.
\end{align}
\end{subequations}
We note that user $2$ suffers zero LSB interference for levels
$n>n_1^*$. Furthermore each receiver can also employ reverse
stripping.
Varying the threshold
$n_2$, \Thmref{FBC-inner-bound-rs}  yields the achievable rate pairs
\begin{subequations}
\begin{align}
R_1 &=2p_1\sum_{n=n_2+1}^{n_1^*}
\entropyhat[n_1^*-n]{3s_1^*2^{-2n}},\\
R_2&=2\sum_{n=1}^{n_2}
\E[S_2]{\entropyhat[n_2-n]{3S_22^{-2n}}}
+2\sum_{n=n_1^*+1}^{\infty}
\E[S_2]{\entropyhat[\infty]{3S_22^{-2n}}}.
\end{align}
\end{subequations}
In Figure~\ref{fig:awgnray}, these rates are tagged ``Inner 1'' and
they are seen to be somewhat worse than an alternate ``Inner 2''
achievable scheme. Note that both schemes employ reverse
stripping. For the ``Inner 2'' scheme, we make the observation
that signals levels $n>n_1^*$ which, following the guidance of the
outer bound assignment,  are assigned to receiver $2$
actually convey negligible information. In fact, information is
conveyed to user $2$ on these signal levels only when the user $1$
SNR exceeds $60~dB$, which is rare since user $2$ has an average SNR
of $30$~dB. However assigning these levels to user $2$ penalizes the
reverse stripping mechanism of receiver $1$. In particular, by not
transmitting on these levels, user $1$ will face no LSB interference
and thus will be able to obtain the ``Inner 2'' rates
\begin{subequations}
\begin{align}
R_1 &=2p_1\sum_{n=n_2+1}^{n_1^*}
\entropyhat[\infty]{3s_1^*2^{-2n}},\\
R_2&=2\sum_{n=1}^{n_2}
\E[S_2]{\entropyhat[n_2-n]{3S_22^{-2n}}}.
\end{align}
\end{subequations}
We see in Figure~\ref{fig:awgnray} that this provides receiver $1$ a
half-bit rate increase when the transmitter is largely dedicated to
receiver $1$ with a neglible reduction to receiver $2$.

Note however that the edge for ``Inner 2'' strategy ceases if
receiver $2$ has sufficiently high average SNR to exploit the high
levels. More generally, we observe that following the guidance of the
outer bound may not maximize the BES rates. In fact, one could directly
optimize the BES achievable rates. However, even the simple example of
Figure~\ref{fig:awgnray} shows there are non-obvious tradeoffs between
the allocation of signal levels and the  rate improvements afforded by
enhancements to reverse stripping.

\section{Fading Gaussian BC: A Constant Gap Result}\label{sect:const-gap}
With knowledge of the distributions $\ccdf{S_i}{s}$,
Theorems~\thmref{FBC-outer} and~\thmref{FBC-inner-bound-rs} can be
used for direct calculation of outer and inner bounds to the fading BC
capacity region.  Now we show that the achievable rates of
\Thmref{FBC-inner-bound} are within a constant gap of the outer bound
rates of \Thmref{FBC-outer} for all channel state distributions.

The key idea in matching up the inner and outer bounds is a
quantization of the channel state analagous to the signal levels of
layered erasure channel. In particular, we let $\gamma>0$ denote a
constant to be determined later and define the channel states
\begin{align}
\gamma_n&= \gamma 2^{2(n-1)},& n&=1,2,\ldots,\\
\Gamma_n&=[\gamma_n,\gamma_{n+1}),&  n&=1,2,\ldots.
\end{align}
In the absence of specific distributions for
$S_1$ and $S_2$, we enlarge the \Thmref{FBC-outer} outer bound  by
enhancing and discretizing the channel gains $S_1$ and $S_2$.
We define the enhanced channels $\Senh_1$ and $\Senh_2$
by the complementary CDFs
\begin{equation}\eqnlabel{enh-outer-bound}
\ccdf{\Senh_i}{s}=\begin{cases}
1 & s\le \gamma\\
\ccdf{S_i}{\gamma_n} & s\in \Gamma_n,\qquad
n=1,2,\ldots\\
\end{cases}
\end{equation}
Since $\ccdf{\Senh_i}{s}\ge \ccdf{S_i}{s}$, the outer bound of
\Thmref{FBC-outer} correpsonding to the fading Gaussian BC
$(\Senh_1,\Senh_2)$ contains the \Thmref{FBC-outer} outer bound region
of the fading Gaussian BC
$(S_1,S_2)$.  For this enhanced channel, the extreme points of the rate region are
given by
\begin{subequations}
\eqnlabel{R-outer2}
\begin{align}
\Renh_1&=\log e\int_{\Iset_1(\w)}
\ccdf{\Senh_1}{s}\frac{1}{1+s}\,ds\\
\Renh_2 &= \log e\int_{\Iset_2(\w)} \ccdf{\Senh_2}{s}\frac{1}{1+s}\,ds.
\end{align}
\end{subequations}
with $\Iset_i(\w)$ given by \eqnref{Iset-cts} with $S_i$ replaced by
the enhanced channel $\Senh_i$.
However, because $\ccdf{\Senh_i}{s}$ is constant over each interval
$\Gamma_n$,
\begin{equation}
\Iset_i(\w) \subseteq [0,\gamma)\cup
\bigcup_{n\in\Nset_i(\w)} \Gamma_n
\end{equation}
such that
\begin{subequations}\eqnlabel{Nset-cts}
\begin{align}
\Nset_1(\w)
&=\set{n\ge1|\ccdf{S_1}{\gamma_n}>\w\ccdf{S_2}{\gamma_{n}}},\\
\Nset_2(\w)&=\set{n\ge 1|\ccdf{S_1}{\gamma_n}\le
  \w\ccdf{S_2}{\gamma_{n}}}.
\end{align}
\end{subequations}
Since $\ccdf{\Senh_i}{s}=1$ for $s\le\gamma$, and is constant over each
interval $\Gamma_n$, it follows from \eqnref{R-outer2}
that
\begin{align}\eqnlabel{Renh-bound}
\Renh_i&\le \log(1+\gamma)
+\sum_{n\in\Nset_i(\w)}\ccdf{S_i}{\gamma_n}[C(n+1)-C(n)]
\end{align}
where
\begin{equation}
C(n)=\log(1+\gamma_n)
\end{equation}
is the capacity of the point-to-point channel with SNR $\gamma_n$.
We note that
\begin{equation}
\Cinc{n} :=C(n+1)-C(n)
\end{equation}
is the incremental capacity associated with the channel improving from
state $\gamma_n$ to state $\gamma_{n+1}$. Note that $\Cinc{n} \le 2$
and approaches this upper bound when $n \rightarrow \infty$.  This
roughly matches the binary expansion model of
Section~\ref{sect:layered-erasure}, in that going from state $n$ to
state $n+1$ in the complex channel yields approximately one additional
bit each for the in-phase and quadrature channels. Applying the bound
$\Cinc{n}\le 2$ to \eqnref{Renh-bound}, we obtain the outer
bound
\begin{align}\eqnlabel{FBC-outer2}
R_i&\le \log(1+\gamma) +2\sum_{n\in\Nset_i(\w)}\ccdf{S_i}{\gamma_n}.
\end{align}

For the purpose of a constant gap result, we start with BES achievable
rate without reverse stripping. From \eqnref{raten-1}, user $i$ can
communicate reliably on level $n\in\hat{\Nset}_i$ at rate
\begin{align}\eqnlabel{raten-1a}
\rate{n}&\ge 2\int_{\gamma_n}^\infty
\pdf{S_i}{s}[1-H(\epshat_0(a_n(s)))]\,ds\\
&=2\ccdf{S_i}{\gamma_n}
-2\levelgap{n}.
\end{align}
where
\begin{align}
\levelgap{n}
&= \sum_{k=n}^\infty
\int_{\Gamma_k} \pdf{S_i}{s} H(\epshat_0(a_n(s)))\,ds.
\end{align}
Note that $a_n(s)$ is an increasing function of $s$ and $\epshat_0(a)$
is a decreasing function of $a$. Thus $\epshat_0(a_n(s))$ is a
decreasing function of $s$.
Since $\epshat_0(a_n(s))\le 1/2$,
$H(\epshat_0(a_n(s)))$ is a decreasing function of $s$. Thus for
$s\in\Gamma_k$,
\begin{equation}
H(\epshat_0(a_n(s)))\le H(\epshat_0(a_n(\gamma_k)))
=H(\epshat_0(3\gamma_{k-n})).
\end{equation}
This implies
\begin{align}
\levelgap{n}
&\le \sum_{k=n}^\infty
H(\epshat_0(3\gamma_{k-n}))
\prob{S_i\in  \Gamma_k}\\
&\le \sum_{m=0}^\infty
H(\epshat_0(3\gamma_m))
\prob{S_i\in \Gamma_{m+n}}.
\end{align}
With the assignment of signal levels $n\in\hat{\Nset}_i$, user $i$
can achieve a rate
\begin{equation}
R_i=2\sum_{n\in\hat{\Nset}_i}\rate{n} \ge 2\sum_{n\in\hat{\Nset}_i}\ccdf{S_i}{\gamma_n}-\levelsum,
\end{equation}
where
\begin{equation}
\levelsum=\sum_{n\in\hat{\Nset}_i}\levelgap{n}
\end{equation}
satisfies the upper bound
\begin{align}
\levelsum\le \sum_{n=1}^\infty \levelgap{n} &= \sum_{n=1}^\infty\sum_{m=0}^\infty
H(\epshat_0(3\gamma_{m}))
\prob{S_i\in \Gamma_{m+n}}\\
&= \sum_{m=0}^\infty H(\epshat_0(3\gamma_m))
\ccdf{S_i}{\gamma_{m+1}}\\
&\le \sum_{m=0}^\infty H(\epshat_0(3\gamma_m)).\eqnlabel{raten-last}
\end{align}
In terms of the channel state set
$\set{\gamma_n}$, we obtain  the inner bound
\begin{align}\eqnlabel{FBC-inner2}
R_i&\ge  2\sum_{n\in\hat{\Nset}_i}\ccdf{S_i}{\gamma_n}
-2\sum_{m=0}^\infty H(\epshat_0(3\gamma_m)).
\end{align}
In comparing the inner bound \eqnref{FBC-inner2} to the outer bound
\eqnref{FBC-outer2}, the rate gap for user $i$ is
given by
\begin{equation}\eqnlabel{general-gap}
\Delta_i=\log(1+\gamma)+2\sum_{m=0}^\infty
H(\epshat_0(3\gamma_m))
\end{equation}
As $\gamma_m=\gamma2^{2(m-1))}$, this bound can be  minimized by
choosing $\gamma=5.65$, yielding $\Delta_i\le 6.386$. This is a
universal gap that holds for all channel state distributions. This
is significantly larger than the gaps observed in the examples of
Section~\ref{sect:awgn-examples}  because this is a {\em worst-case}
bound on the gap over {\em all} fading distributions.  We note that
our specific examples showed the reverse stripping scheme can
deliver a 2-3 bit improvement in bit rates. In the worst case,
however, reverse stripping is unlikely to help because the
transmission scheme may end up assigning every other level to a
user.

On the other hand, there is still some potential on improving even
the worst-case gap. We observe that in the quantization of channel
states in 6~dB steps, the outer bound is loosened by assuming a
receiver always obtains the best channel in each interval while the
inner bound is tightened by the opposite assumption that a receiver
always gets the worst channel. The impact of quantization appears to
be on the order of two bits. Thus we conjecture that the actual
worst-case gap is considerably smaller than six bits. Nevertheless,
the 6.386 bit gap does demonstrate that the BES signaling has the
right asymptotic behavior in high SNR.

\section{Conclusion}
This work derives the first constant gap result for the capacity region of
the AWGN fading broadcast channel with channel state information
known at the receivers only.  Our calculations show that the rate
gap can then be bounded by $6.386$ bits/s/Hz universally for all
fading distributions. To obtain this conclusion, we derive a new
outer bound and use a  simple achievability strategy, both of which are
motivated by the analysis of an approximating layered erasure
broadcast channel. We conjecture that more careful analysis and more
sophisticated achievability schemes will shrink this gap
considerably.

\appendix
\section{Proofs}\label{app:proofs}
\begin{Proof}{Lemma~\ref{lem:qbit-fade}}
\begin{letterate}
\item As $N$ is a deterministic function of $X^N$,
\begin{align}
I(X^q;X^N|V)&=I(X^q;X^N,N|V)\\
&=I(X^q;N|V)+I(X^q;X^N|V,N).
\end{align}
Since the channel state $N$ is independent of $V$ and $X^q$,
$I(X^q;N|V)=0$. Thus
\begin{align}
I(X^q;X^N|V)&=I(X^q;X^N|V,N)\\
&=H(X^N|V,N),
\end{align}
since $H(X^N|V,N,X^q)=0$.

\item Since $V$ and $X^n$ are independent of $N$,
\begin{align}
H(X^N|V,N)&=\sum_{n=1}^q \pmf{N}{n}H(X^n|V,N=n)\\
&=\sum_{n=1}^q \pmf{N}{n} H(X^n|V).
\end{align}
Applying the chain rule and  reversing the order
of summation then yields
\begin{align}
H(X^N|V,N)
&=\sum_{n=1}^q \sum_{j=1}^n \pmf{N}{n}H(X_j|X^{j-1},V)
\\
&=\sum_{j=1}^q\sum_{n=j}^q \pmf{N}{n}
H(X_j|X^{j-1},V)
\\
&=\sum_{j=1}^q\ccdf{N}{j}H(X_j|X^{j-1},V).
\end{align}
\item As $N$ is a deterministic function of $X^N$,
\begin{align}
I(V;X^N) &= I(V;X^N,N)\\
&=I(V;N)+I(V;X^N|N).
\end{align}
Since $V$ is independent of the channel state
$N$, $I(V;N)=0$ and
\begin{align}
I(V;X^N) &=I(V;X^N|N)\\
&= H(X^N|N)-H(X^N|V,N).
\end{align}
Applying the result of part~(a) to $H(X^N|N)$ (with a trivial $V$) and
also to $H(X^N|V,N)$  yields
\begin{align}
I(V;X^N)&=\sum_{j=1}^q\ccdf{N}{j}
\bigl[H(X_j|X^{j-1})
-H(X_j|V,X^{j-1})\bigl]\\
&=\sum_{j=1}^q\ccdf{N}{j} I(V;X_j|X^{j-1}).
\end{align}
\end{letterate}
\end{Proof}


\begin{Proof}{Lemma~\ref{lem:Ideriv}}
From \eqnref{Rstar-cts-start}, independence of $X$ and $\tilde{S}_1$
and independence of $V$ and $S_2$ imply
\begin{align}
R^*(\w)&\le\max_{V,X} I(X;Y_1|V,\tilde{S}_1) + \omega I(V;Y_2|S_2) \\
&= \max_{V,X} h(Y_1|V,\tilde{S}_1) - h(Z_1) + \omega \left [
  h(Y_2|S_2) - h(Y_2|V,S_2) \right ] \\
&= \max_{V,X} h(Y_1|V,\Stil_1)-\w h(Y_2|V_1,S_2)
+\w h(Y_2|S_2)-h(Z),\eqnlabel{FBC-Rstar}
\end{align}
In terms of $\Ys$, we can write
 \begin{align}
h(Y_1|V,\Stil_1)&=\int_0^\infty
\pdf{\Stil_1}{s}h(\sqrt{s}X+Z|V,\Stil=s)\,ds,\\
&=\int_0^\infty \pdf{\Stil_1}{s}h(\sqrt{s}X+Z|V)\,ds,\\
&=\int_0^\infty \pdf{\Stil_1}{s}h(\Ys|V)\,ds.\eqnlabel{hY1givenVS1}
\end{align}
Similarly,
\begin{align}h(Y_2|V,S_2)&=\int_0^\infty \pdf{S_2}{s} h(\Ys|V)\,ds
\eqnlabel{hY2givenVS2}\end{align}
and
\begin{align}
h(Y_2|S_2)&=\int_0^\infty \pdf{S_2}{s} h(\sqrt{s}X+Z_2|S_2=s)\,ds\\
&\le \int_0^\infty \pdf{S_2}{s}\frac{1}{2}\log [2\pi e(s+1)]\,ds\\
&= h(Z)+C_e(S_2).\eqnlabel{hY2givenS2}
\end{align}

Applying \eqnref{hY1givenVS1}, \eqnref{hY2givenVS2} and
\eqnref{hY2givenS2} to \eqnref{FBC-Rstar}, we obtain
\begin{align}
R^*(\w)&\le \max_{V,X}
\int_0^\infty \pdf{\w}{s} h(\Ys|V)\,ds -
(1-\w)h(Z) +\w C_e(S_2)\eqnlabel{Rstar-pdf}
\end{align}
where
$\pdf{\w}{s}=\pdf{\Stil_1}{s}-\w\pdf{S_2}{s}$.
Since $\pdf{\w}{s}=-d\ccdf{\w}{s}/ds$,  integration by parts can be used
to show that for any $g(s)$ satisfying
$\ccdf{\w}{\infty}g(\infty)=0$ that
\begin{equation}
\int_0^\infty \pdf{\w}{s}g(s)\,ds =\ccdf{\w}{0}g(0)+\int_0^\infty
\ccdf{\w}{s}g'(s)\,ds.
\end{equation}
Note that $\ccdf{\w}{0}=1-\w$ and that $g(s)=h(\Ys|V)$ implies
$g(0)=\eval{h(\Ys|V)}_{s=0}=h(Z)$.  This implies
\begin{equation}
\int_0^\infty \pdf{\w}{s}h(\Ys|V)=(1-\w)h(Z)+\int_0^\infty
\ccdf{\w}{s}h'(\Ys|V)\,ds \eqnlabel{hYbyparts},
\end{equation}
where $h'(\Ys|v)$ denotes the derivative of $h(\Ys|V)$ with respect to $s$.
It follows from \eqnref{Rstar-pdf} and \eqnref{hYbyparts} that
\begin{align}
R^*(\w)&\le \max_{V,X} \int_0^\infty
\ccdf{\w}{s}h'(\Ys|V)\,ds+\w C_e(S_2).
\end{align}
The claim follows since $I(X;\Ys|V)= h(\Ys|V)-h(Z)$,
implies $I'(X;\Ys|V)=h'(\Ys|V)$.
\end{Proof}

\begin{Proof}{Lemma~\ref{lem:crossover-d}}
From \eqnref{noncrossprob-d}, 
\begin{align}
\pc{\xtil^n} 
&=\prob{2^{-n}(-1-2^{n-d}\Util_n)\le\Ztil< 2^{-n}(1-2^{n-d}\Util_n)}\\
&= \prob{-1+U\le Z_n< 1+U},
\end{align}
where $U=-2^{n-d}U_n$ is a continuous uniform $(-2^{-d},2^{-d})$
random variable and 
$Z_n=2^n\Ztil$ has standard deviation
$\sigma_n=2^n/\sqrt{3s}=1/\sqrt{a_n(s)}$.
Since $U$ and $Z_n$ are  independent,
\begin{align}
\pc{\xtil^n} &=2^{d-1}\int_{-2^{-d}}^{2^{-d}}
\prob{-1+u\le Z_n< 1+u}\,du\\
&=2^{d-1}\int_{-2^{-d}}^{2^{-d}}\bracket{1-Q\parfrac{1+u}{\sigma_n}-Q\parfrac{1-u}{\sigma_n}}\,du\\
&=1-\sigma_n 2^d\int_{(1-2^{-d})/\sigma_n}^{(1+2^{-d})/\sigma_n}
Q(v)\,dv.
\end{align}
Noting that $G(x)=\int Q(x)\,dx$ and that $\sigma_n=1/\sqrt{a_n(s)}$, we
obtain
\begin{equation}
1-\pc{\xtil^n}= \frac{G(\sqrt{a_n(s)}(1+2^{-d})) -
  G(\sqrt{a_n(s)}(1-2^{-d}))}{\sqrt{a_n(s)}2^{-d}}=\epsilon_d(a_n(s)).
\end{equation}
Since the conditional probability $\pc{\xtil^n}$ does not depend on
  $\xtil^n$, it follows that the error (i.e. crossover) probability of
  this detector for 
bit $\Xtil_n$ under channel state $S=s$ with LSB interference at depth
$d$ satisfies
\begin{equation}\eqnlabel{pen1}
\pcross_{n,d}(s)\le \prob{\Xhat_n\neq\Xtil_n}\le 1-\pc{\xtil^n}=
\epsilon_d(a_n(s)).
\end{equation}
The claim follows by using \eqnref{binary-detector} to guess bits when
the channel state is weak.
\end{Proof}
\bibliographystyle{unsrt}
\bibliography{ry-it}
\end{document}